%% file: idolip.tex
\documentclass[pre,twocolumn,showpacs,showkeys,superscriptaddress,preprintnumbers,floatfix]{revtex4-1}

\usepackage{etex}
\usepackage{ifpdf}
\usepackage{hyperref}
\usepackage{dcolumn}
\usepackage{url}
\usepackage{amsmath}
\usepackage{amscd}
\usepackage{amsfonts}
\usepackage{amssymb}
\usepackage{bm}   
\usepackage{bbm}
\usepackage{verbatim}
\usepackage{stmaryrd}
\usepackage{amsthm}
\usepackage{xcolor}
\usepackage{setspace}
\usepackage{braket}

\usepackage{tikz}
\usetikzlibrary{arrows}
\usetikzlibrary{automata}
\usetikzlibrary{decorations.pathreplacing}
\usetikzlibrary{positioning}
\usetikzlibrary{plotmarks}
\usetikzlibrary{calc}
\usetikzlibrary{patterns}
\usetikzlibrary{external}
\usepackage{pgfplots}
\pgfplotsset{compat=newest}

\usepackage{graphicx}

\theoremstyle{plain}    
\theoremstyle{plain}    
\theoremstyle{plain}    
\theoremstyle{plain}    
\theoremstyle{plain}    
\theoremstyle{plain}    
\theoremstyle{plain}    
\theoremstyle{plain}    
\theoremstyle{plain}    
\theoremstyle{plain}    
\theoremstyle{plain}    
\theoremstyle{plain}

\input{cmechabbrev}

\renewcommand{\H}{\operatorname{H}}
\renewcommand{\I}{\operatorname{I}}

\newcommand{\kB}{k_\text{B}}  

\begin{document}


\title{Above and Beyond the Landauer Bound:\\
Thermodynamics of Modularity}

\author{Alexander B. Boyd}
\email{abboyd@ucdavis.edu}
\affiliation{Complexity Sciences Center and Physics Department,
University of California at Davis, One Shields Avenue, Davis, CA 95616}

\author{Dibyendu Mandal}
\email{dibyendu.mandal@berkeley.edu}
\affiliation{Department of Physics, University of California, Berkeley, CA
94720, U.S.A.}

\author{James P. Crutchfield}
\email{chaos@ucdavis.edu}
\affiliation{Complexity Sciences Center and Physics Department,
University of California at Davis, One Shields Avenue, Davis, CA 95616}

\date{\today}
\bibliographystyle{unsrt}

\begin{abstract}
Information processing typically occurs via the composition of modular units,
such as universal logic gates. The benefit of modular information processing,
in contrast to globally integrated information processing, is that complex
global computations are more easily and flexibly implemented via a series of
simpler, localized information processing operations which only control and
change local degrees of freedom. We show that, despite these benefits, there
are unavoidable thermodynamic costs to modularity---costs that arise directly
from the operation of localized processing and that go beyond Landauer's
dissipation bound for erasing information. Integrated computations can achieve
Landauer's bound, however,  when they globally coordinate the control of all of
an information reservoir's degrees of freedom. Unfortunately, global
correlations among the information-bearing degrees of freedom are easily lost
by modular implementations. This is costly since such
correlations are a thermodynamic fuel. We quantify the minimum irretrievable dissipation of modular computations in terms of the difference between the
change in global nonequilibrium free energy, which captures these global
correlations, and the local (marginal) change in nonequilibrium free energy,
which bounds modular work production. This \emph{modularity dissipation} is
proportional to the amount of additional work required to perform the
computational task modularly. It has immediate consequences for physically
embedded transducers, known as information ratchets. We show how to circumvent
modularity dissipation by designing internal ratchet states that capture the
global correlations and patterns in the ratchet's information reservoir.
Designed in this way, information ratchets match the optimum thermodynamic
efficiency of globally integrated computations. Moreover, for ratchets that
extract work from a structured pattern, minimized modularity dissipation means
their hidden states must be predictive of their input and, for ratchets that
generate a structured pattern, this means that hidden states are retrodictive.
\end{abstract}


\preprint{Santa Fe Institute Working Paper 17-08-XXX}
\preprint{arxiv.org:1708.XXXXX [cond-mat.stat-mech]}

\maketitle



\setstretch{1.1}
\section{Introduction}

Physically embedded information processing operates via thermodynamic
transformations of the supporting material substrate. The thermodynamics is
best exemplified by Landauer's principle: erasing one bit of stored information
at temperature $T$ must be accompanied by the dissipation of at least $\kB T
\ln{2}$ amount of heat \cite{Land61a} into the substrate. While the Landauer
cost is only time-asymptotic and not yet the most significant energy demand in
everyday computations---in our cell phones, tablets, laptops, and cloud
computing---there is a clear trend and desire to increase thermodynamic
efficiency. Digital technology is expected, for example, to reach the vicinity
of the Landauer cost in the near future; a trend accelerating with
now-promising quantum computers. This seeming inevitability forces us to ask if
the Landauer bound can be achieved for more complex information processing
tasks than writing or erasing a single bit of information.


In today's massive computational tasks, in which vast arrays of bits are
processed in sequence and in parallel, a task is often broken into modular
components to add flexibility and comprehensibility to hardware and software
design. This holds far beyond the arenas of today's digital computing. Rather
than tailoring processors to do only the task specified, there is great benefit
in modularly deploying elementary, but universal functional components---e.g.,
NAND, NOR, and perhaps Fredkin \cite{Fred82b} logic gates, biological neurons
\cite{Riek99}, or similar units appropriate to other domains
\cite{Alon07a}---that can be linked together into circuits which perform any
functional operation. This leads naturally to hierarchical design, perhaps
across many organizational levels. In these ways, the principle of modularity
reduces the challenges of designing, monitoring, and diagnosing efficient
processing considerably \cite{Ulri94, Chen16}. Designing each modular component
of a complex computation to be efficient is vastly simpler than designing and
optimizing the whole. Even biological evolution seems to have commandeered
prior innovations, remapping and reconnecting modular functional units to form
new organizations and new organisms of increasing survivability \cite{Mayn98a}.

There is, however, a potential thermodynamic cost to modular information
processing. For concreteness, recall the stochastic computing paradigm in which
an input (a sequence of symbols) is sampled from a given probability
distribution and the symbols are correlated to each other. In this setting, a
modularly designed computation processes only the \textit{local} component of
the input, ignoring the latter's \textit{global} structure. This inherent
locality necessarily leads to irretrievable loss of the global correlations
during computing. Since such correlations are a thermal resource
\cite{Boyd16c,Lloy89}, their loss implies an energy cost---a thermodynamic
\emph{modularity dissipation}. Employing stochastic thermodynamics and
information theory, we show how modularity dissipation arises by deriving an
exact expression for dissipation in a generic localized information processing
operation. We emphasize that this dissipation is above and beyond the Landauer
bound for losses in the operation of single logical gates. It arises solely
from the modular architecture of complex computations. One immediate
consequence is that the additional dissipation requires investing additional
work to drive computation forward.


In general, to minimize work invested in performing a computation, we must
leverage the global correlations in a system's environment. Globally integrated
computations can achieve the minimum dissipation by simultaneous control of the
whole system, manipulating the joint system-environment Hamiltonian to follow
the desired joint distribution. Not only is this level of control difficult to
implement physically, but designing the required protocol poses a considerable
computational challenge in itself, with so many degrees of freedom and a
potentially complex state space. Genetic algorithm methods have been proposed,
though, for approximating the optimum \cite{Stor96}. Tellingly, they can find
unusual solutions that break conventional symmetries and take advantage of the
correlations between the many different components of the entire system
\cite{Lohn05, Koza92a}. However, as we will show, it is possible to rationally
design local information processors that, by accounting for these correlations,
minimized modularity dissipation.

The following shows how to design optimal modular computational schemes such
that useful global correlations are not lost, but stored in the structure of
the computing mechanism. Since the global correlations are not lost in these
optimal schemes, the net processing can be thermodynamically reversible
(dissipationless). Utilizing the tools of information theory and computational
mechanics---Shannon information measures and optimal hidden Markov
generators---we identify the informational system structures that can
mitigate and even nullify the potential thermodynamic cost of modular
computation.


A brief tour of our main results will help orient the reader. It can even serve
as a complete, but approximate description for the approach and technical
details, should this be sufficient for the reader's interests.

Section \ref{sec:GlobalvLocal} considers the thermodynamics of a composite
information reservoir, in which only a subsystem is amenable to external
control. In effect, this is our model of a localized thermodynamic operation.
We assume that the information reservoir is coupled to an ideal heat bath, as a
source of randomness and energy. Thus, external control of the information
reservoir yields random Markovian dynamics over the informational states, heat
flows into the heat bath, and work investment from the controller. Statistical
correlations may exist between the controlled and uncontrolled subsystems,
either due to initial or boundary conditions or due to an operation's history.

To highlight the information-theoretic origin of the dissipation and to
minimize the energetic aspects, we assume that the informational states have
equal internal (free) energies. Appealing to stochastic thermodynamics and
information theory, we then show that the minimum irretrievable
\emph{modularity dissipation} over the duration of an operation due to the
locality of control is proportional to the reduction in mutual information
between the controlled and uncontrolled subsystems; see
Eq.~(\ref{eq:Irretrievable}). We deliberately refer to ``operation'' here
instead of ``computation'' since the result holds whether the desired task is
interpreted as computation or not. The result holds so long as free-energy
uniformity is satisfied at all times, a condition natural in computation and
other information processing settings.


Section \ref{sec:InfoTransducers} applies this analysis to \emph{information
engines}, an active subfield within the thermodynamics of computation in which
information effectively acts as the fuel for driving physically embedded
information processing~\cite{Mand012a,Boyd15a,Merh15a,Boyd16d,Boyd16c}. The
particular implementations of interest---\emph{information ratchets}---process
an input symbol string by interacting with each symbol in order, sequentially
transforming it into an output symbol string, as shown in
Fig.~\ref{fig:FullRatchet}. This kind of \emph{information
transduction}~\cite{Barn13a,Boyd15a} is information processing in a very
general sense: with a properly designed finite-state control, the devices can
implement a universal Turing machine \cite{Broo89a}. Since information engines
rely on localized information processing, reading in and manipulating one
symbol at a time in their original design~\cite{Mand012a}, the measure of
irretrievable dissipation applies directly. The exact expression for the
modularity dissipation is given in Eq.~(\ref{eq:InfoEngines}).


Sections \ref{sec:Extractors} and \ref{sec:Generators} specialize information
transducers further to the cases of pattern extractors and pattern generators.
Section \ref{sec:Extractors}'s \emph{pattern extractors} use structure in their
environment to produce work and \emph{pattern generators} use stored work to
create structure from an unstructured environment. The irreversible relaxation
of correlations in information transduction can then be curbed by intelligently
designing these computational processes. While there are not yet general
principles for designing implementations for arbitrary computations, the
measure of modularity dissipation that we develop in the following shows how to
construct energy-efficient extractors and generators. For example, efficient
extractors consume complex patterns and turn them into sequences of independent
and identically distributed (IID) symbols.

We show that extractor transducers whose states are predictive of their inputs are optimal, with zero minimal modularity dissipation. This makes immediate
intuitive sense since, by design, such transducers can anticipate the next
input and adapt accordingly. This observation also emphasizes the principle
that thermodynamic agents should requisitely match the structural complexity of
their environment to leverage those informational correlations as a
thermodynamic fuel~\cite{Boyd16d}. We illustrate this result in the case of the
Golden Mean pattern in Fig.~\ref{fig:GoldenMean}.

Conversely, Section \ref{sec:Generators} shows that when generating patterns
from unstructured IID inputs, transducers whose states are retrodictive of
their output are most efficient---i.e., have minimal modularity dissipation.
This is also intuitively appealing in that pattern generation may be viewed as
the time reversal of pattern extraction. Since predictive transducers are
efficient for pattern extraction, retrodictive transducers are expected to be
efficient pattern generators; see Fig.~\ref{fig:RetrodictiveGenerator}. This
also allows one to appreciate that pattern generators previously thought to be
asymptotically efficient are actually quite dissipative~\cite{Garn15}. Taken
altogether, these results provide guideposts for designing efficient, modular,
and complex information processors---guideposts that go substantially beyond
Landauer's principle for localized processing.


\section{Global versus Localized Processing}
\label{sec:GlobalvLocal}


\newcommand{\ISystem}{\mathcal{Z}}
\newcommand{\IInter}{\ISystem^{\text{int}}}
\newcommand{\IStat}{\ISystem^{\text{stat}}}
\newcommand{\RVSystem}{Z}
\newcommand{\RVInter}{\RVSystem^{\text{i}}}
\newcommand{\RVStat}{\RVSystem^{\text{s}}}
\newcommand{\rvsystem}{z}
\newcommand{\rvinter}{\rvsystem^{\text{i}}}
\newcommand{\rvstat}{\rvsystem^{\text{s}}}

If a physical system, denote it $\ISystem$, stores information as it
behaves, it acts as an information reservoir. Then, a wide range of
physically-embedded computational processes can be achieved by connecting
$\ISystem$ to an ideal heat bath at temperature $T$ and externally controlling
the system's physical parameters, its Hamiltonian. Coupling with the heat bath
allows for physical phase-space compression and expansion, which are necessary
for useful computations and which account for the work investment and heat
dissipation dictated by Landauer's bound. However, the bound is only achievable
when the external control is precisely designed to harness the changes in
phase-space. This may not be possible for modular computations. The modularity
here implies that control is localized and potentially ignorant of global
correlations in $\ISystem$. This leads to uncontrolled changes in phase-space.

\begin{figure}[tbp]
\centering
\includegraphics[width=.9\columnwidth]{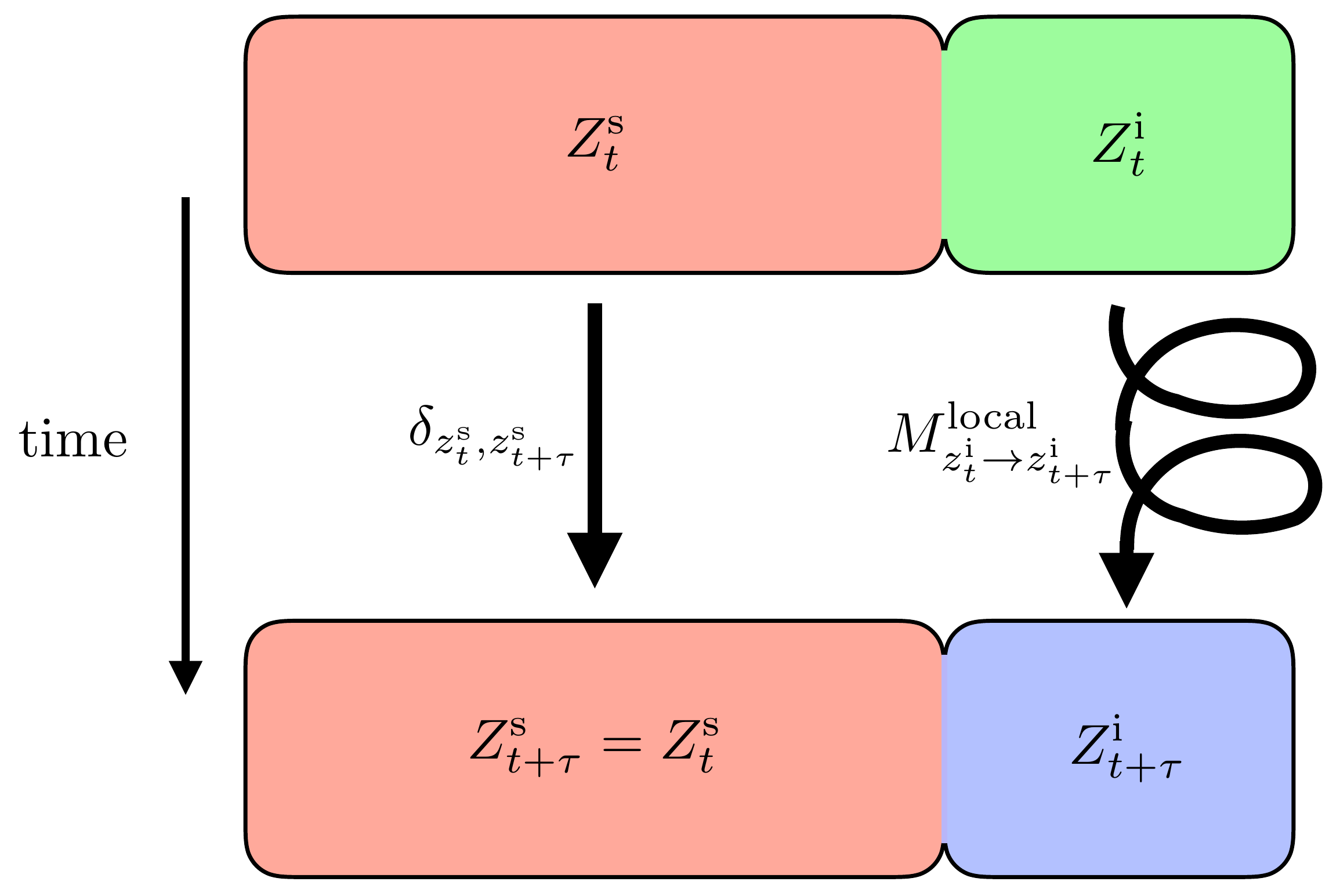}
\caption{Local computations operate on only a subset $\IInter$ of the entire
	information reservoir $\ISystem = \IInter \otimes \IStat$. The Markov
	channel that describes the global dynamic is the product of a local
	operation with the identity operation: $M^\text{global}_{(\rvinter_t,
	\rvstat_t) \rightarrow (\rvinter_{t+\tau}, \rvstat_{t+\tau})} =
	M^\text{local}_{\rvinter_t \rightarrow \rvinter_{t+\tau}}
	\delta_{\rvstat_t,\rvstat_{t+\tau}}$, such that the stationary
	noninteracting portion $\RVStat$ of the information reservoir remains
	invariant, but the interacting portion $\RVInter$ changes.
	}
\label{fig:LocalComputation}
\end{figure}

Most computational processes unfold via a sequence of local operations that
update only a portion of the system's informational state. A single step in
such a process can be conveniently described by breaking the whole
informational system $\ISystem$ into two constituents: the informational states
$\IInter$ that are controlled and evolving and the informational states
$\IStat$ that are not part of the local operation on $\IInter$. We call
$\IInter$ the \emph{interacting} subsystem and $\IStat$ the \emph{stationary}
subsystem. As shown in Fig. \ref{fig:LocalComputation}, the dynamic over the
joint state space $\ISystem = \IInter \otimes \IStat$ is the product of the
identity over the stationary subsystem and a local Markov channel over the
interacting subsystem. The informational states of the noninteracting
stationary subsystem $\IStat$ are fixed over the immediate computational task,
since this information should be preserved for use in later computational steps.

Such classical computations are described by a global Markov channel over the
joint state space:
\begin{align}
&M^\text{global}_{\rvinter_t, \rvstat_t \rightarrow \rvinter_{t+\tau},
\rvstat_{t+\tau}}
  \nonumber \\
  & \quad = \Pr(\RVInter_{t+\tau} \! = \! \rvinter_{t+\tau},
  \RVStat_{t+\tau} \! = \! \rvstat_{t+\tau} | \RVInter_{t}
  \! = \! \rvinter_{t}, \RVStat_t \! = \! \rvstat_t)
  \label{eq:TrPr}
  ,
\end{align}
where $\RVSystem_t = \RVInter_t \otimes \RVStat_t$ and
$\RVSystem_{t+\tau} = \RVInter_{t+\tau} \otimes \RVStat_{t+\tau}$ are the
random variables for the informational state of the joint system
before and after the computation, with $\RVInter$ describing the
$\IInter$ subspace and $\RVStat$ the $\IStat$ subspace, respectively. 
(Lowercase variables denote values their associated random variables realize.) The righthand side of
Eq.~(\ref{eq:TrPr}) gives the transition probability over the time interval
$(t, t + \tau)$ from joint state $(\rvinter_t, \rvstat_t)$ to state
$(\rvinter_{t+\tau}, \rvstat_{t+\tau})$. The fact that
$\IStat$ is fixed means that the global dynamic can be
expressed as the product of a local Markov computation on
$\IInter$ with the identity over $\IStat$:
\begin{align}
M^\text{global}_{(\rvinter_t, \rvstat_t)
	\rightarrow (\rvinter_{t+\tau}, \rvstat_{t+\tau})}
	= M^\text{local}_{\rvinter_t \rightarrow
	\rvinter_{t+\tau}} \delta_{\rvstat_t,\rvstat_{t+\tau}}
  ~,
\label{eq:MPLocal}
\end{align}
where the local Markov computation is the conditional marginal distribution:
\begin{align}
 M^\text{local}_{\rvinter_t \rightarrow \rvinter_{t+\tau}}
   = \Pr(\RVInter_{t+\tau}= \rvinter_{t+\tau}| \RVInter_t = \rvinter_t)
   ~.
\end{align}

When the processor is in contact with a heat bath at temperature $T$, the average entropy production $\langle \Sigma_{t \rightarrow t + \tau} \rangle$ of the universe over the time interval $(t,t+\tau)$ can be expressed in terms of the work done minus the change in nonequilibrium free energy $F^\text{neq}$:
\begin{align*}
 \langle \Sigma_{t \rightarrow t +\tau} \rangle=\frac{ \langle W_{t \rightarrow t +\tau} \rangle -(F^\text{neq}_{t+\tau}-F^\text{neq}_{t})}{T}
   ~.
\end{align*}
In turn, the nonequilibrium free energy $F^\text{neq}_t$ at any time $t$ can be
expressed as the weighted average of the internal (free) energy $U_{z}$ of the
joint informational states minus the uncertainty in those states:
\begin{align*}
F^\text{neq}_t
  & = \sum_{\rvsystem} \Pr(\RVSystem=\rvsystem) U_{\rvsystem}
  - \kB T \ln 2 \H[\RVSystem_t]
  ~.
\end{align*}
Here, $\H[Z]$ is the Shannon information of the random variable $Z$ that
realizes the state of the joint system $\ISystem$ \cite{Parr15a}. When the
information bearing degrees of freedom support an information reservoir, where
all states $\rvsystem$ and $\rvsystem'$ have the same internal energy $U_{z} =
U_{z'}$, the entropy production reduces to the work minus a change in Shannon
information of the information-bearing degrees of freedom:
\begin{align}
\langle \Sigma_{t \rightarrow t +\tau} \rangle
  = \frac{\langle W_{t \rightarrow t +\tau} \rangle}{T}
  \! + \! \kB \ln 2(\H[\RVSystem_{t+\tau}] \! - \! \H[\RVSystem_t])
  ~.
\label{eq:Landauer}
\end{align}
Essentially, this is an expression of a generalized Landauer Principle: entropy
increase guarantees that work production is bounded by the change in Shannon
entropy of the informational variables \cite{Land61a}.

In particular, for a globally integrated quasistatic operation, where all degrees of freedom
are controlled simultaneously as discussed in App. \ref{app:Quasistatic Markov
Channels}, there is zero entropy production. And, the globally integrated work done on the
system achieves the theoretical minimum:
\begin{align*}
\langle W_{t \rightarrow t+\tau}^\text{global} \rangle_\text{min}
  = - \kB T \ln 2 ( \H[\RVSystem_{t+\tau}] - \H[\RVSystem_t] )
  ~.
\end{align*}
The process is reversible since the change in system Shannon entropy balances
the change in the reservoir's physical entropy due to heat dissipation. (Since
the internal energy is uniform, the system cannot store the work and must
dissipate it as heat to the surrounding environment.) This may not be the case
for a generic modular operation. 


There are two consequences of the locality of control. First, since $\RVStat$
is kept fixed---that is, $\RVStat_t = \RVStat_{t+\tau}$---the change in
uncertainty $\H[\RVInter_{t+\tau},\RVStat_{t+\tau}]-\H[\RVInter_t,\RVStat_t]$
of the joint informational variables during the operation, which is the
second term in lefthand side of Eq. (\ref{eq:Landauer}), simplifies to:
\begin{align*}
\H[\RVSystem_{t+\tau}]-\H[\RVSystem_t]
  = \H[\RVInter_{t+\tau},\RVStat_t]
  - \H[\RVInter_t,\RVStat_t]
  ~.
\end{align*}  
Second, since we operate locally on $\RVInter$, with no knowledge of $\RVStat$,
then the required work is bounded by the generalized Landauer Principle
corresponding to the marginal distribution over $\RVInter$; see
Eq.~(\ref{eq:MPLocal}). In other words, in absence of any control over the
noninteracting subsystem $\RVStat$, which remains stationary over the local
computation on $\RVInter$, the minimum work performed on $\RVInter$ is given
by:
\begin{align*}
\langle W_{t \rightarrow t + \tau} \rangle
  & \geq \langle W^\text{local}_{t \rightarrow t +\tau} \rangle_\text{min} \\
  & = \kB T \ln 2 ( \H[\RVInter_t]-H[\RVInter_{t+\tau})
  ~.
\end{align*}
This bound is achievable through a sequence of quasistatic and instantaneous
protocols, described in App. \ref{app:Quasistatic Markov Channels}.

Combining the last two relations with the expression for entropy production in
Eq.~(\ref{eq:Landauer}) gives the \emph{modularity dissipation}
$\Sigma^\text{mod}$, which is the minimum irretrievable dissipation of a
modular computation that comes from local interactions:
\begin{align}
\frac{\langle \Sigma^\text{mod}_{t \rightarrow t +\tau}
\rangle_\text{min}}{\kB \ln 2}
  & = \frac{ \langle W^\text{local}_{t \rightarrow t +\tau}
  \rangle_\text{min}}{k_B T \ln2}
  + \H[\RVInter_{t+\tau},\RVStat_t]-\H[\RVInter_t,\RVStat_t] \nonumber \\
  &  = \I[\RVInter_t;\RVStat_t]-I[\RVInter_{t+\tau};\RVStat_t]
  ~,
\label{eq:Irretrievable}
\end{align}
where $\I[X;Y]$ is the mutual information between the random variables $X$ and $Y$. 

This is our central result: there is a thermodynamic cost above and beyond the
Landauer bound for modular operations. It is a thermodynamic cost arising from
a computation's \emph{implementation architecture}. Specifically, the minimum
entropy production is proportional to the minimum additional work that must be
done to execute a computation modularly:
\begin{align*}
\langle W^\text{local}_{t \rightarrow t +\tau} \rangle_\text{min}
  - \langle W^\text{global}_{t \rightarrow t +\tau} \rangle_\text{min}
  = T \langle \Sigma^\text{mod}_{t \rightarrow t +\tau} \rangle_\text{min}
  ~.
\end{align*}
The following draws out the implications.

\begin{figure}[tbp]
\centering
\includegraphics[width=.9\columnwidth]{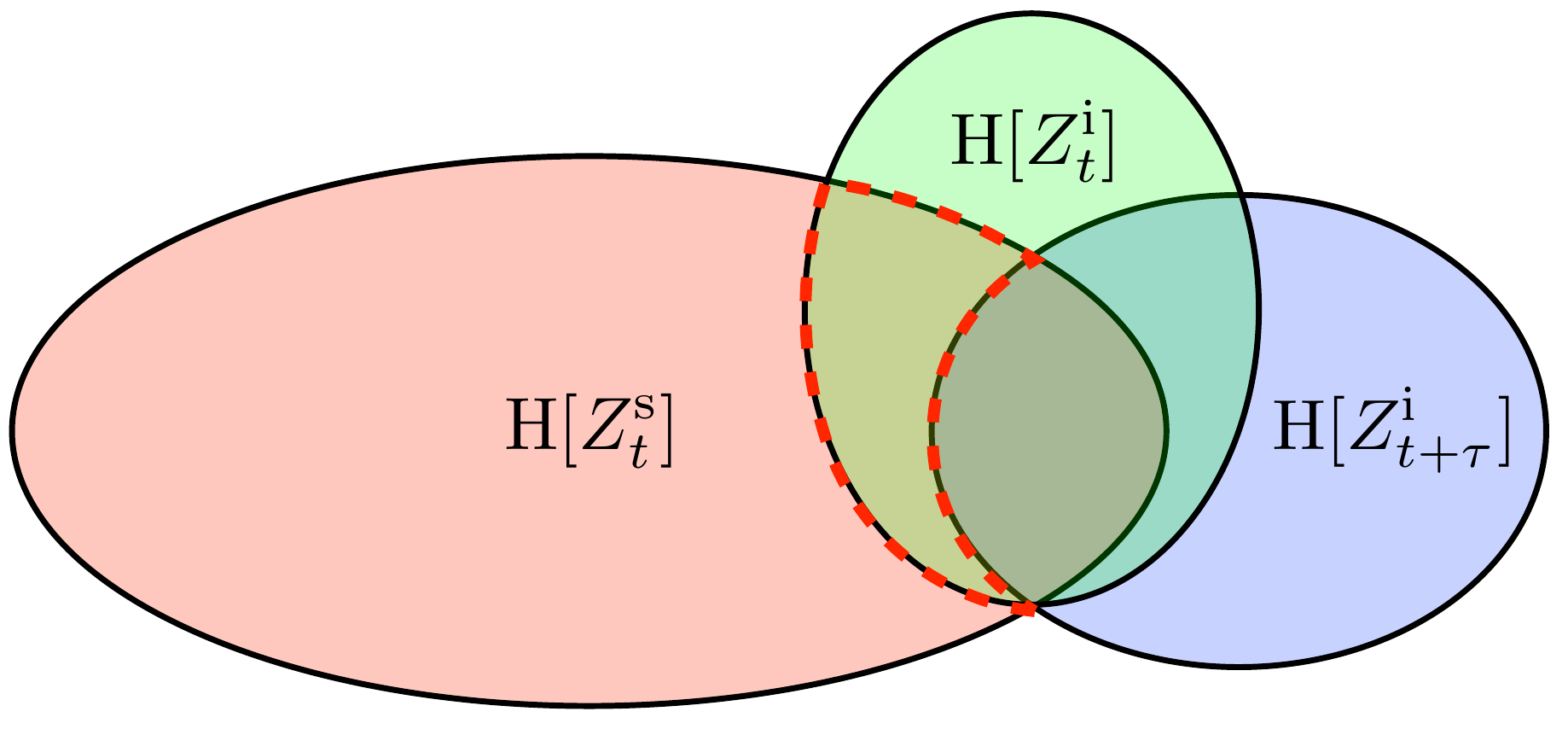}
\caption{Information diagram for a local computation: Information atoms of the
	noninteracting subsystem $\H[\RVStat_t]$ (red ellipse), the interacting
	subsystem before the computation $\H[\RVInter_t]$ (green circle), and the
	interacting subsystem after the computation $\H[\RVInter_{t+\tau}]$ (blue
	circle). The initial state of the interacting subsystem shields the final
	state from the noninteracting subsystem; graphically the blue and red
	ellipses only overlap within the green ellipse. The modularity
	dissipation is proportional to the difference between information atoms
	$\I[\RVInter_t;\RVStat_t]$ and $\I[\RVInter_{t+\tau};\RVStat_t]$.
	Due to statistical shielding, it simplifies to the information atom
	$\I[\RVInter_t;\RVStat_t|\RVInter_{t+\tau}]$, highlighted by a red dashed
	outline.
	}
\label{fig:IrretrievableDissipation}
\end{figure}


Using the fact that the local operation $M^\text{local}$ ignores $\RVStat$, we
see that the joint distribution over all three variables $\RVInter_t, \RVStat_t,$
and $\RVInter_{t + \tau}$ can be simplified to:
\begin{align*}
\Pr(\RVInter_{t+\tau} & = \rvinter_{t+\tau},\RVInter_t=\rvinter_t,\RVStat_t=\rvstat_t) \\
  & = \Pr(\RVInter_{t+\tau}=\rvinter_{t+\tau}|\RVInter_t=\rvinter_t)\Pr(\RVInter_t=\rvinter_t,\RVStat_t=\rvstat_t)
  ~.
\end{align*}
Thus, $\RVInter_t$ \emph{shields} $\RVInter_{t+\tau}$ from $\RVStat_t$.  A
consequence is that the mutual information between $\RVInter_{t + \tau}$ and
$\RVStat_t$ conditioned on $\RVInter_t$ vanishes. This is shown in
Fig.~\ref{fig:IrretrievableDissipation} via an information diagram.
Figure~\ref{fig:IrretrievableDissipation} also shows that the modularity
dissipation, highlighted by a dashed red outline, can be re-expressed as the
mutual information between the noninteracting stationary system $\RVStat$ and
the interacting system $\RVInter$ before the computation that is not shared
with $\RVInter$ after the computation:
\begin{align}
\langle \Sigma^\text{mod}_{t \rightarrow t +\tau} \rangle_\text{min}
  = \kB \ln 2 \I[\RVInter_t;\RVStat_t | \RVInter_{t +\tau}]
  ~.
\label{eq:ConditionalIrretrievable}
\end{align}
This is our second main result. The conditional mutual information on the right
bounds how much entropy is produced when performing a local computation.
It quantifies the irreversibility of information processing.

We close this section by noting that Eq.~(\ref{eq:Irretrievable})'s bound is
analogous to the expression for the minimum work required for data
representation, with $\RVInter_t$ being the work medium, $\RVInter_{t+\tau}$ the
work extraction device, and $\RVStat_t$ the data representation device
\cite{Stil17a}.

The following unpacks the implications of Eqs.~\eqref{eq:Irretrievable}
and~\eqref{eq:ConditionalIrretrievable} for information
transducers---information processing architectures in which the processor
sequentially takes one input symbol at a time and performs localized
computation on it, much as a Turing machine operates.

\section{Information Transducers: Localized processors}
\label{sec:InfoTransducers}

\begin{figure}[tbp]
\centering
\includegraphics[width=.9\columnwidth]{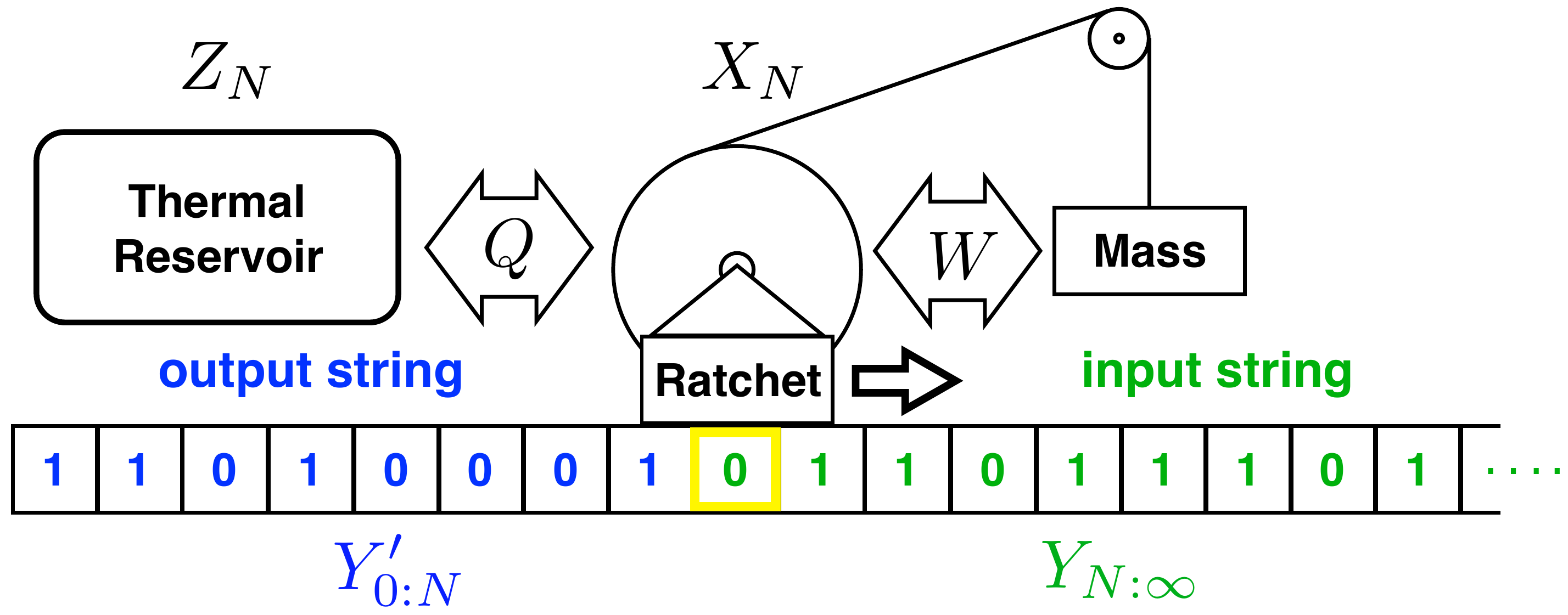}
\caption{Information ratchet consists of three interacting reservoirs---work,
	heat, and information. The work reservoir is depicted as gravitational mass
	suspended by a pulley. The thermal reservoir keeps the entire system
	thermalized to temperature $T$. At time $N\tau$ the information reservoir
	consists of (i) a string of symbols $\mathbf{Y}_N = Y'_0 Y'_1 \ldots
	Y'_{N-1} Y_N Y_{N+1} \ldots$, each cell storing an element from the same
	alphabet $\mathcal{Y}$ and (ii) the ratchet's internal state $X_N$. The
	ratchet moves unidirectionally along the string, exchanging energy between
	the heat and the work reservoirs. The ratchet reads the value of a single
	cell (highlighted in yellow) at a given time from the input string (green,
	right), interacts with it, and writes a symbol to the cell in the output
	string (blue, left) of the information reservoir. Overall, the ratchet
	transduces the input string $Y_{0:\infty} =Y_0 Y_1 \ldots$ into an output
	string $Y'_{0:\infty} =Y'_0 Y'_1 \ldots$.
	(Reprinted from Ref. \protect\cite{Boyd15a} with permission.)
	}
\label{fig:FullRatchet}
\end{figure}

Information ratchets \cite{Lu14a,Boyd15a} are thermodynamic implementations of
information transducers \cite{Barn13a} that sequentially transform an input
symbol string, described by the chain of random variables $Y_{0:\infty}=Y_0 Y_1
Y_2, \ldots$, into an output symbol string, described by the chain of random
variables $Y'_{0:\infty}= Y'_0 Y'_1 Y'_2, \ldots$. The ratchet traverses the
input symbol string unidirectionally, processing each symbol in turn to yield
the output sequence. As shown in Fig. \ref{fig:FullRatchet}, at time $t=N \tau$
the information reservoir is described by the joint distribution over the
ratchet state $X_N$ and the symbol string $\mathbf{Y}_N = Y_{0:N}'
Y_{N:\infty}$, the concatenation of the first $N$ symbols of the output string
and the remaining symbols of the input string. (This differs slightly from
previous treatments \cite{Boyd16c} in which only the symbol string is the
information reservoir. The information processing and energetics are the same,
however.) Including the ratchet state in present definition of the information
reservoir allows us to directly determine the modularity dissipation of
information transduction.

Going from time $t=N\tau$ to $t+\tau=(N+1)\tau$ preserves the state of
the current output history $Y_{0:N}'$ and the input future, excluding the $N$th
symbol $Y_{N+1:\infty}$, while changing the $N$th input symbol $Y_N$ to the
$N$th output symbol $Y_N'$ and the ratchet from its current state $X_N$ to
its next $X_{N+1}$. In terms of the previous section, this means the
noninteracting stationary subsystem $\IStat$ is the entire semi-infinite
symbol string \emph{without} the $N$th symbol:
\begin{align}
\RVStat_t = (Y_{N+1:\infty}, Y'_{0:N})
  ~.
\end{align}
The ratchet and the $N$th symbol constitute the interacting subsystem
$\IInter$ so that, over the time interval $(t,t+\tau)$, only two variables
change:
\begin{align}
\RVInter_t = (X_N, Y_N)
\end{align}
and
\begin{align}
\RVInter_{t+\tau} = (X_{N+1}, Y_N')
  ~.
\end{align}

Despite the fact that only a small portion of the system changes on each time
step, the physical device is able to perform a wide variety of physical and
logical operations. Ignoring the probabilistic processing aspects, Turing
showed that a properly designed (very) finite-state transducer can compute any
input-output mapping \cite{Turi37a} \footnote{Space limitations here do not
allow a full digression on possible implementations. Suffice it to say that for
unidirectional tape reading, the ratchet state requires a storage register or
an auxiliary internal working tape as portrayed in Fig. 3 of Ref.
\cite{Crut92c}.}. Such machines, even those with as few as two internal states
and a sufficiently large symbol alphabet \cite{Shan56c} or with as few as a
dozen states but operating on a binary-symbol strings, are \emph{universal} in
that sense \cite{Mins67}.

Information ratchets---physically embedded, probabilistic Turing machines---are
able to facilitate energy transfer between a thermal reservoir at temperature
$T$ and a work reservoir by processing information in symbol strings. In
particular, they can function as an eraser by using work to create structure in
the output string \cite{Mand012a,Boyd15a} or act as an engine by using the
structure in the input to turn thermal energy into useful work energy
\cite{Boyd15a}. They are also capable of much more, including detecting,
adapting to, and synchronizing to environment correlations
\cite{Boyd16e,Boyd16d} and correcting errors \cite{Boyd16c}.

Information transducers are a novel form of information processor from a
different perspective, that of communication theory's channels \cite{Barn13a}.
They are memoryful channels that map input stochastic processes to output
processes using internal states which allow them to store information about the
past of both the input and the output. With sufficient hidden states, as just
noted from the view of computation theory, information transducers are Turing
complete and so able to perform any computation on the information reservoir
\cite{Stra15a}. Similarly, the physical steps that implement a transducer as
an information ratchet involve a series of modular local computations.

The ratchet operates by interacting with one symbol at a time in sequence, as
shown in Fig. \ref{fig:FullRatchet}. The $N$th symbol, highlighted in yellow
to indicate that it is the interacting symbol, is changed from the input $Y_N$
to output $Y'_N$ over time interval $(N\tau,(N+1)\tau)$. The ratchet and
interaction symbol change together according to the local Markov channel over
the ratchet-symbol state space:
\begin{align*}
M^\text{local}_{(x,y) \rightarrow (x',y')}
  = \Pr(X_{N+1} \! = \! x',Y_N' \! = \! y'|X_N \! = \! x,Y_N \! = \! y)
  .
\end{align*}
This determines how the ratchet transduces inputs to outputs \cite{Boyd15a}.

Each of these localized operations keeps the remaining noninteracting symbols
in the information reservoir fixed. If the ratchet only has energetic control
of the degrees of freedom it manipulates, then, as discussed in the previous
section and App. \ref{app:Quasistatic Markov Channels}, the ratchet's work
production in the $N$th time step is bounded by the change in uncertainty of
the ratchet state and interaction symbol:
\begin{align}
\langle W^\text{local}_N \rangle_\text{min}
   \! =  \! \kB T \ln2 (\H[X_N,Y_N] \! - \! \H[X_{N+1},Y'_N])
  .
\label{eq:LocalBound}
\end{align}
This bound has been recognized in previous investigations of information
ratchets \cite{Mand012a, Merh17a}. Here, we make a key, but important and
compatible observation: If we relax the condition of local control of energies
to allow for global control of all symbols simultaneously, then it is possible
to extract more work.

That is, foregoing localized operations---abandoning modularity---allows for
(and acknowledges the possibility of) globally integrated interactions. Then,
we can account for the change in Shannon information of the information
reservoir---the ratchet and the entire symbol string. This yields a looser
upper bound on work production that holds for both modular and globally
integrated information processing. Assuming that all information reservoir
configurations have the same free energies, the change in the nonequilibrium
free energy during one step of a ratchet's computation is proportional to the
global change in Shannon entropy:
\begin{align*}
\Delta F^\text{neq}_{N \tau \rightarrow (N+t)\tau}
  \! = \! \kB T \ln 2 (\H[X_N,\mathbf{Y}_N]
  \! - \! \H[X_{N+1},\mathbf{Y}_{N+1}])
  .
\end{align*}

Recalling the definition of entropy production $\langle \Sigma \rangle =
\langle W \rangle -\Delta F^\text{neq}$ reminds us that for entropy to
increase, the minimum work investment must match the change in free energy:
\begin{align}
\langle & W^\text{global}_N \rangle_\text{min} \nonumber \\
  & \quad = \kB T \ln 2 (\H[X_N, \mathbf{Y}_N]
  - \H[X_{N+1}, \mathbf{Y}_{N+1}])
  ~.
\label{eq:NonLocalBound}
\end{align}
This is the work production that can be achieved through globally integrated quasistatic information processing. And, in turn, it can be used to bound the asymptotic work production in terms of the entropy rates of the input and output processes
\cite{Boyd15a}:
\begin{align}
\lim_{N\rightarrow \infty} \langle W_N \rangle
  \geq \kB T \ln 2 (\hmu-\hmu')
  ~.
\end{align}  
This is known as the \emph{Information Processing Second Law} (IPSL).

Reference \cite{Boyd16d} already showed that this bound is not necessarily
achievable by information ratchets. This is due to ratchets operating locally.
The local bound on work production of modular implementations in Eq. (\ref{eq:LocalBound}) is less
than or equal to the global bound on integrated implementations in Eq. (\ref{eq:NonLocalBound}), since the
local bound ignores correlations between the interacting system $\IInter$ and
noninteracting elements of the symbol string in $\IStat$. Critically, though,
if we design the ratchet such that its states store the relevant correlations
in the symbol string, then we can achieve the global bounds. This was hinted at
in the fact that the gap between the work done by a ratchet and the global
bound can be closed by designing a ratchet that matches the input process'
structure \cite{Boyd16c}. However, comparing the two bounds now allows us to be
more precise.

The difference between the two bounds represents the amount of additional work
that could have been performed by a ratchet, if it was not modular and limited
to local interactions. If the computational device is globally integrated, with
full access to all correlations between the information bearing degrees of
freedom, then all of the nonequilibrium free energy can be converted to work,
zeroing out the entropy production. Thus, the minimum entropy production for a
modular transducer (or information ratchet) at the $N$th time step can be
expressed in terms of the difference between Eq.  (\ref{eq:LocalBound}) and the
entropic bounds in Eq. (\ref{eq:NonLocalBound}):
\begin{align}
\label{eq:InfoEngines}
\frac{\langle \Sigma^\text{mod}_N \rangle_\text{min}}{\kB \ln 2}
  & =\frac{\langle W^\text{local}_N\rangle_\text{min}
  - \Delta F^\text{neq}_{N \tau \rightarrow (N+1)\tau}}{k_B T \ln 2} \\
  & =  \I[Y_{N+1:\infty}, Y'_{0:N};X_N,  Y_N] \nonumber \\
  & \qquad - \I[Y_{N+1:\infty}, Y'_{0:N};X_{N+1}, Y'_N] \nonumber \\
  & = \I[Y_{N+1:\infty}, Y'_{0:N};X_N,  Y_N|X_{N+1}, Y'_N]
  \label{eq:IrretrievableTransducer}
  ~.
\end{align}
This can also be derived directly by substituting our interacting variables
$(X_N, Y_N) = \RVInter_t$ and $(X_{N+1},Y'_N) = \RVInter_{t+\tau}$ and
stationary variables $(Y_{N+1:\infty},Y'_{0:N}) = \RVStat$ into the expression
for the modularity dissipation in Eqs. (\ref{eq:Irretrievable}) and
(\ref{eq:ConditionalIrretrievable}) in Sec. \ref{sec:GlobalvLocal}. Even if the
energy levels are controlled so slowly that entropic bounds are reached, Eq.
(\ref{eq:IrretrievableTransducer}) quantifies the amount of lost correlations
that cannot be recovered. And, this leads to the entropy production and
irreversibility of the transducing ratchet. This has immediate consequences
that limit the most thermodynamically efficient information processors.

\begin{figure*}[tbp]
\centering
\includegraphics[width=2\columnwidth]{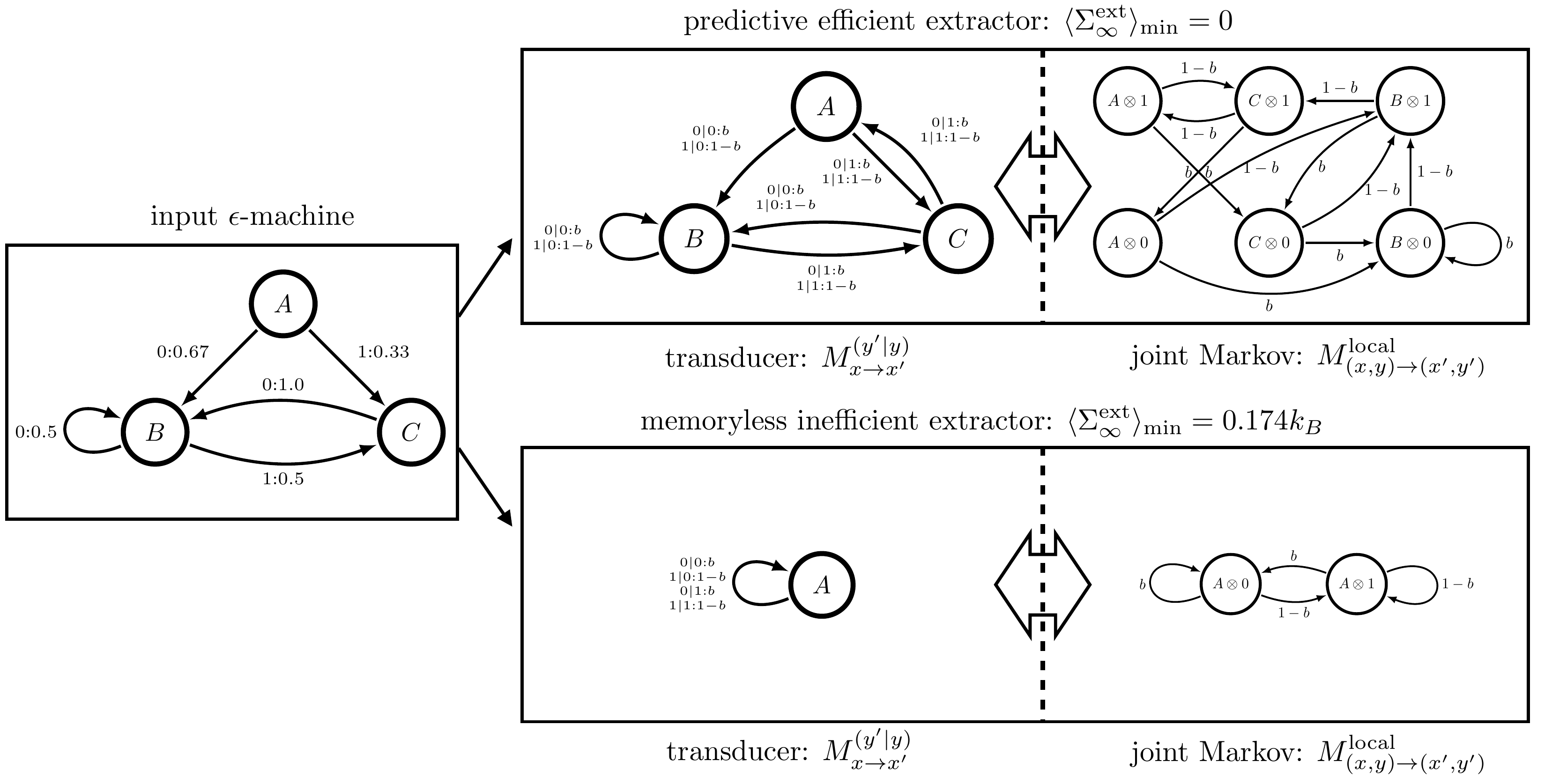}
\caption{Multiple ways to transform the Golden Mean Process input, whose
	\eM\ generator is shown in the far left box, into a sequence of
	uncorrelated symbols. The \eM\ is a Mealy hidden Markov model that produces
	outputs along the edges, with $y:p$ denoting that the edge emits symbol $y$
	and is taken with probability $p$. (Top row) Ratchet whose internal states
	match the \eM\ states and so it is able to minimize dissipation---$\langle
	\Sigma^\text{ext}_\infty \rangle_\text{min}=0$---by making transitions
	such that the ratchet's states are synchronized to the \eM's states. The
	transducer representation to the left shows how the states remain
	synchronized: its edges are labeled $y'|y:p$, which means that if the input
	was $y$, then with probability $p$ the edge is taken and it outputs $y'$.
	The joint Markov representation on the right depicts the corresponding
	physical dynamic over the joint state space of the ratchet and the
	interaction symbol. The label $p$ along an edge from the state $x\otimes y$
	to $x' \otimes y'$ specifies the probability of transitioning between those
	states according to the local Markov channel $M^\text{local}_{(x,y)
	\rightarrow (x', y')}=p$. (Bottom row) In contrast to the efficient
	predictive ratchet, the memoryless ratchet shown is inefficient, since it's
	memory cannot store the predictive information within the input \eM, much
	less synchronize to it.
	}
\label{fig:GoldenMean}
\end{figure*}

While previous bounds, such as the IPSL, demonstrated that information in the
symbol string can be used as a thermal fuel
\cite{Mand012a,Boyd15a}---leveraging structure in the inputs symbols to turn
thermal energy into useful work---they largely ignore the structure of
information ratchet states $X_N$. The transducer's hidden states, which can
naturally store information about the past, are critical to taking advantage of
structured inputs. Until now, we only used informational bounds to predict
transient costs to information processing \cite{Garn15, Boyd16e}. With the
expression for the modularity dissipation of information ratchets in Eq.
(\ref{eq:IrretrievableTransducer}), however, we now have bounds that apply to
the ratchet's asymptotic functioning. In short, this provides the key tool for
designing thermodynamically efficient transducers. We will now show that it has
immediate implications for pattern generation and pattern extraction.

\section{Predictive Extractors}
\label{sec:Extractors}

A \emph{pattern extractor} is a transducer that takes in a structured process
$\Pr(Y_{0:\infty})$, with correlations among the symbols, and maps it to a
series of independent identically distributed (IID), uncorrelated output
symbols. An output symbol can be distributed however we wish individually, but
each must be distributed with an identical distribution and independently from
all others. The result is that the joint distribution of the output process
symbols is the product of the individual marginals:
\begin{align}
\Pr(Y_{0:\infty}') = \prod_{i=0}^{\infty} \Pr(Y_{i}')
  ~.
\end{align}
If implemented efficiently, this device can use temporal correlations in the
input as a thermal resource to produce work. The modularity dissipation of
an extractor $\langle  \Sigma_N^\text{ext} \rangle_\text{min}$ can be
simplified by noting that the output symbols are uncorrelated with any other
variable and, thus, fall out of the mutual information terms:
\begin{align*}
\frac{\langle \Sigma_N^\text{ext} \rangle_\text{min}}{k_B \ln 2}
  =  \I[Y_{N+1:\infty};X_N, Y_N] - \I[Y_{N+1:\infty};X_{N+1}]
  ~.
\end{align*}
Minimizing this irreversibility, as shown in App. \ref{app:Dissipation of
Predictive and Retrodictive Transducers}, leads directly to a fascinating
conclusion that relates thermodynamics to prediction: the states of maximally
thermodynamically efficient extractors are optimally predictive of the input
process.

To take full advantage of the temporal structure of an input process,
the ratchet's states $X_N$ must be able to predict the future of the input
$Y_{N:\infty}$ from the input past $Y_{0:N}$. Thus, the ratchet shields the
input past from the output future such that there is no information shared
between the past and future which is not captured by the ratchet's states:
\begin{align}
\I[Y_{N:\infty};Y_{0:N}|X_N] = 0
  ~.
\label{eq:Predictive}
\end{align}
Additionally, transducers cannot anticipate the future of the inputs beyond
their correlations with past inputs \cite{Barn13a}. This means that there is no
information shared between the ratchet and the input future when conditioned on
the input past:
\begin{align}
\I[Y_{N:\infty};X_N|Y_{0:N}] = 0
  ~.
\label{eq:NonAnticipatory}
\end{align}
Together, Eqs. (\ref{eq:Predictive}) and (\ref{eq:NonAnticipatory}) are
equivalent to both the state $X_N$ being predictive and the modularity
dissipation vanishing $\langle \Sigma_N^\text{ext} \rangle_\text{min}=0$. The
efficiency of predictive ratchets suggests that predictive generators, such as
the $\epsilon$-machine \cite{Crut01a}, are useful in designing efficient
information engines that can leverage temporal structure in an environment.

For example, consider an input string that is structured according to the
Golden Mean Process, which consists of binary strings in which $1$'s always
occur in isolation, surrounded by $0$'s. Figure \ref{fig:GoldenMean} gives
two examples of ratchets, described by different local Markov channels
$M^\text{local}_{(x,y) \rightarrow (x',y')}$, that each map the Golden Mean
Process to a biased coin. The input process' \eM, shown in left box, provides a
template for how to design a thermodynamically efficient local Markov channel,
since its states are predictive of the process. The Markov channel is a
transducer \cite{Boyd15a}:
\begin{align}
M^{(y'|y)}_{x \rightarrow x'}\equiv M^\text{local}_{(x,y) \rightarrow (x',y')}
  ~.
\end{align}
By designing transducer states that stay synchronized to the states of the
process' \eM, we can minimize the modularity dissipation to zero. For
example, the efficient transducer shown in Fig. \ref{fig:GoldenMean} has almost
the same topology as the Golden Mean \eM, with an added transition between
states $C$ and $A$ corresponding to a disallowed word in the input. This
transducer is able to harness all structure in the input because it
synchronizes to the input process and so is able to optimally predict the next
input.

The efficient ratchet shown in Fig. \ref{fig:GoldenMean} (top row) comes from a
general method for constructing an optimal extractor given the input's \eM. The
\eM\ is represented by a Mealy hidden Markov model (HMM) with the
symbol-labeled state-transition matrices:
\begin{align}
T^{(y)}_{s \rightarrow s'} = \Pr(Y_N=y,S_{N+1}=s'|S_N=s)
  ~,
\label{eq:HMM}
\end{align}
where $S_N$ is the random variable for the hidden state reading the $N$th input
$Y_N$. If we design the ratchet to have the same state space as the input
process' hidden state space---$\mathcal{X}=\mathcal{S}$---and if we want the
IID output to have bias $\Pr(Y_N=0)=b$, then we set the local Markov channel
over the ratchet and interaction symbol to be:
\begin{align*}
M^\text{local}_{(x,y) \rightarrow (x',y')}
  = \begin{cases}
  b, & \text{if }T^{(y)}_{x \rightarrow x'} \neq 0 \text{ and } y'=0 \\
  1-b, & \text{if }T^{(y)}_{x \rightarrow x'} \neq 0 \text{ and } y'=1
  ~.
\end{cases}
\end{align*}

This channel, combined with normalized transition probabilities, does not
uniquely specify $M^\text{local}$, since there can be forbidden words in the
input that, in turn, lead to \eM\ causal states which always emit a single
symbol. This means that there are joint ratchet-symbol states $(x,y)$ such that
$M_{(x,y) \rightarrow (x',y')}$ is unconstrained. For these states, we may make
any choice of transition probabilities from $(x,y)$, since this state will
never be reached by the combined dynamics of the input and ratchet. The end
result is that, with this design strategy, we construct a ratchet whose memory
stores all information in the input past that is relevant to the future, since
the ratchet remains synchronized to the input's causal states. In this way, it
leverages all temporal order in the input.

By way of contrast, consider a memoryless transducer, such as that shown in
Fig. \ref{fig:GoldenMean} (bottom row). It has only a single state and so
cannot store any information about the input past. As discussed in previous
explorations, ratchets without memory are insensitive to correlations
\cite{Boyd16d, Boyd16c}. This result for stationary input processes is subsumed
by the measure of modularity dissipation. Since there is no uncertainty in
$X_N$, the asymptotic dissipation of memoryless ratchets simplifies to:
\begin{align*}
\langle \Sigma^\text{ext}_\infty \rangle_\text{min}
  & = \lim_{N \rightarrow \infty} \kB \ln 2 ~\I[Y_{N+1: \infty};Y_N] \\
  & = \kB \ln 2 ~ (\H_1-\hmu)
  ~,
\end{align*}
where in the second step we used input stationarity---every symbol has the same
marginal distribution---and so the same single-symbol uncertainty
$\H_1=\H[Y_N]=\H[Y_M]$. Thus, the modularity dissipation of a memoryless
ratchet is proportional to the length-$1$ redundancy $\H_1-\hmu$
\cite{Crut01a}. This is the amount of additional uncertainty that comes from
ignoring temporal correlations. As Fig. \ref{fig:GoldenMean} shows, this means
that a memoryless extractor driven by the Golden Mean Process dissipates
$\langle \Sigma^\text{ext}_\infty \rangle_\text{min} \approx 0.174 \kB$ with
every bit. Despite the fact that both of these ratchets perform the same
computational process---converting the Golden Mean Process into a sequence of
IID symbols---the simpler model requires more energy investment to function,
due to its irreversibility.

\section{Retrodictive Generators}
\label{sec:Generators}

Pattern generators are rather like time-reversed pattern extractors, in that
they take in an uncorrelated input process:
\begin{align}
\Pr(Y_{0:\infty}) = \prod_{i=0}^{\infty} \Pr(Y_{i})
  ~,
\end{align}
and turn it into a structured output process $\Pr(Y_{0:\infty})$ that has
correlations among the symbols. The modularity dissipation of a generator
$\langle \Sigma_N^\text{gen} \rangle_\text{min}$ can also be simplified by
removing the uncorrelated input symbols:
\begin{align*}
\frac{\langle \Sigma_N^\text{gen} \rangle_\text{min}}{\kB \ln2}
  = \I[Y'_{0:N};X_N] - \I[Y'_{0:N};X_{N+1}Y'_N]
  ~.
\end{align*}
Paralleling extractors, App. \ref{app:Dissipation of Predictive and
Retrodictive Transducers} shows that retrodictive ratchets minimize the
modularity dissipation to zero.

Retrodictive generator states carry as little information about the output past
as possible. Since this ratchet generates the output, it must carry all the
information shared between the output past and future. Thus, it shields output
past from output future just as a predictive extractor does for the input
process:
\begin{align*}
\I[Y'_{N:\infty};Y'_{0:N}|X_N] = 0
  ~.
\end{align*}
However, unlike the predictive states, the output future shields the
retrodictive ratchet state from the output past:
\begin{align}
\I[X_N;Y'_{0:N}|Y'_{N:\infty}] = 0
  ~.
\end{align}
These two conditions mean that $X_N$ is retrodictive and imply that the
modularity dissipation vanishes. While we have not established the
equivalence of retrodictiveness and efficiency for pattern generators, as we
have for predictive pattern extractors, there are easy-to-construct examples
demonstrating that diverging from efficient retrodictive implementations leads
to modularity dissipation at every step.

\begin{figure}[tbp]
\centering
\includegraphics[width=\columnwidth]{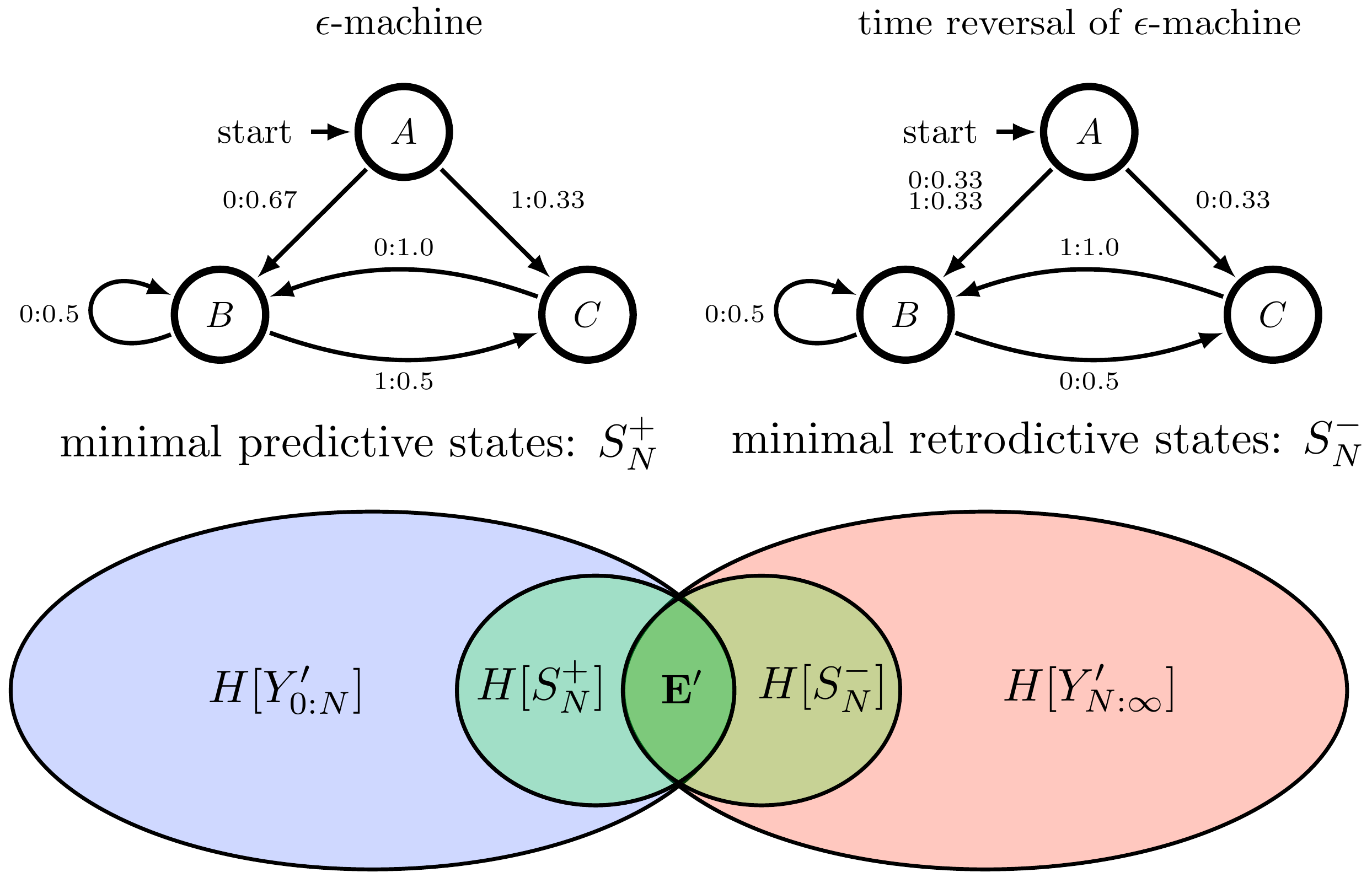}
\caption{Alternate minimal generators of the Golden Mean Process: predictive
	and retrodictive. (Left) The \eM\ has the minimal set of causal states
	$S^+$ required to predictively generate the output process. As a result,
	the uncertainty $\H[S^+_N]$ is contained by the uncertainty $\H[Y'_{0:N}]$
	in the output past. (Right) The time reversal of the reverse-time \eM\ has
	the minimal set of states required to retrodictively generate the output.
	Its states are a function of the output future. Thus, its uncertainty
	$\H[S^-_N]$ is contained by the output future's uncertainty
	$\H[Y'_{N:\infty}]$.
	}
\label{fig:PredictiveVsRetrodictive}
\end{figure}

Consider once again the Golden Mean Process. Figure
\ref{fig:PredictiveVsRetrodictive} shows that there are alternate ways to
generate such a process from a hidden Markov model. The \eM, shown on the left,
is the minimal predictive model, as discussed earlier. It is unifilar, which
means that the current hidden state $S^+_N$ and current output $Y'_N$ uniquely
determine the next hidden state $S^+_{N+1}$ and that once synchronized to the
hidden states one stays synchronized to them by observing only output symbols.
Thus, its states are a function of past outputs. This is corroborated by the
fact that the information atom $\H[S^+_N]$ is contained by the information atom
for the output past $\H[Y'_{0:N}]$. The other hidden Markov model generator
shown in Fig. \ref{fig:PredictiveVsRetrodictive} (right) is the time reversal of
the \eM\ that generates the reverse process. This is much like the \eM, except
that it is retrodictive instead of predictive. The recurrent states $B$ and $C$
are co-unifilar as opposed to unifilar. This means that the next hidden state
$S^-_{N+1}$ and the current output $Y'_N$ uniquely determine the current state
$S^-_N$. The hidden states of this minimal retrodictive model are a function of
the semi-infinite future. And, this can be seen from the fact that the
information atom for $\H[S^-_N]$ is contained by the information atom for the
future $\H[Y'_{N:\infty}]$.

These two different hidden Markov generators both produce the Golden Mean
Process, and they provide a template for constructing ratchets to generate that
process. For a hidden Markov model described by symbol-labeled transition
matrix $\{T^{(y)}\}$, with hidden states in $\mathcal{S}$ as described in Eq.
(\ref{eq:HMM}), the analogous generative ratchet has the same states
$\mathcal{X}=\mathcal{S}$ and is described by the joint Markov local
interaction:
\begin{align*}
M^\text{local}_{(x,y) \rightarrow (x',y')}=T^{(y')}_{x \rightarrow x'}
  ~.
\end{align*}
Such a ratchet effectively ignores the IID input process and obeys the same
informational relationships between the ratchet states and outputs as the
hidden states of hidden Markov model with its outputs.  

\begin{figure*}[tbp]
\centering
\includegraphics[width=2 \columnwidth]{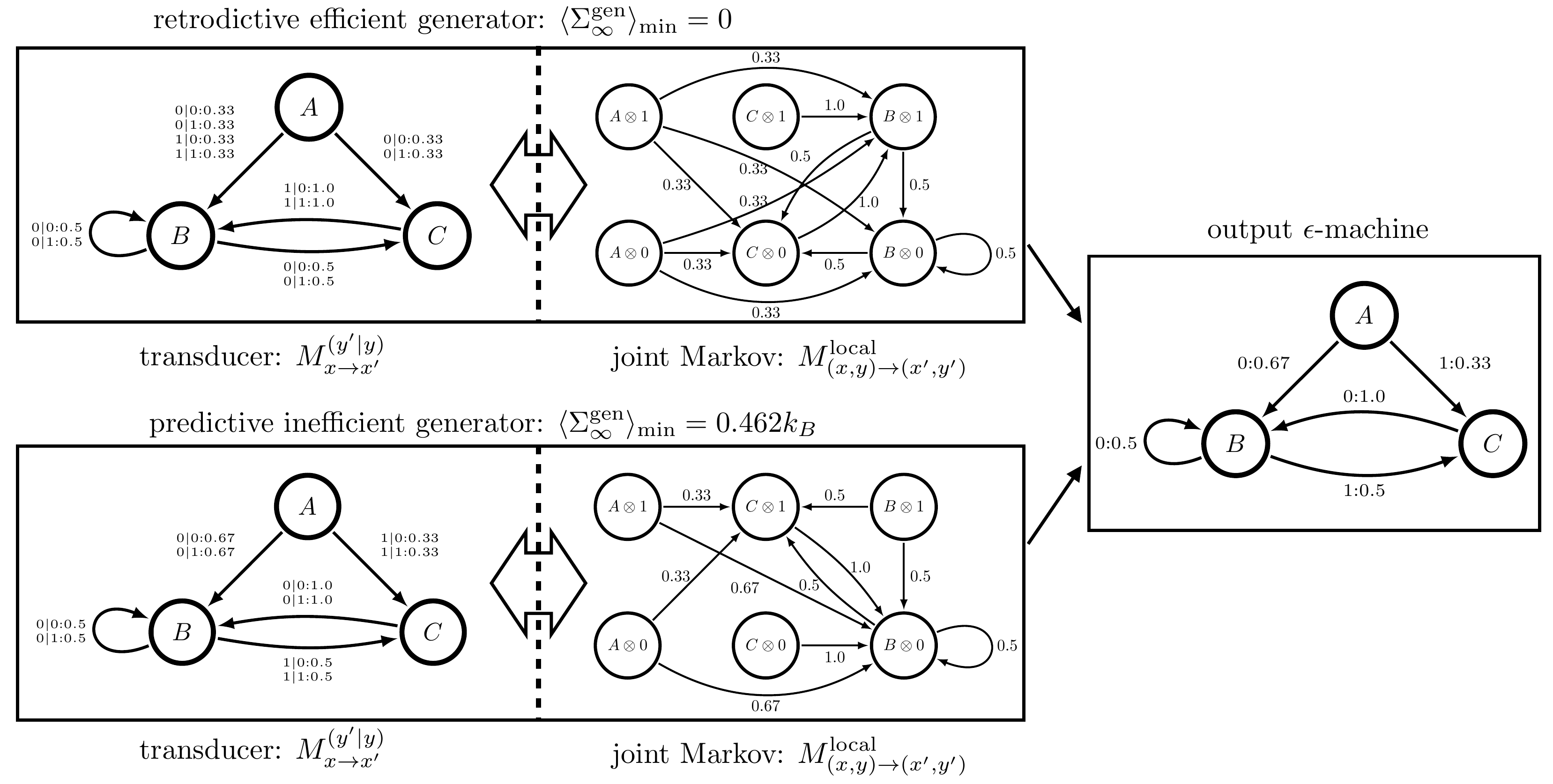}
\caption{Alternative generators of the Golden Mean Process: (Right) The
	process' \eM. (Top row) Optimal generator designed using the topology of
	the minimal retrodictive generator. It is efficient, since it stores as
	little information about the past as possible, while still storing enough
	to generate the output. (Bottom row) The predictive generator stores far
	more information about the past than necessary, since it is based off the
	predictive \eM.  As a result, it is far less efficient. It dissipates at
	least $\tfrac{2}{3} \kB T \ln 2$ extra heat per symbol and requires that
	much more work energy per symbol emitted.
	}
\label{fig:RetrodictiveGenerator}
\end{figure*}

Figure \ref{fig:RetrodictiveGenerator} shows both the transducer and joint
Markov representation of the minimal predictive generator and minimal
retrodictive generator. The retrodictive generator is potentially perfectly
efficient, since the process' minimal modularity dissipation vanishes: $\langle
\Sigma^\text{gen}_N \rangle_\text{min}=0$ for all $N$. However, despite being a
standard tool for generating an output, the predictive \eM\ is necessarily
irreversible and dissipative. The \eM-based ratchet, as shown in Fig.
\ref{fig:RetrodictiveGenerator}(bottom row), approaches an asymptotic dynamic
where the current state $X_N$ stores more than it needs to about the past
output past $Y'_{0:N}$ in order to generate the future $Y'_{N:\infty}$. As a
result, it irretrievably dissipates:
\begin{align*}
\langle \Sigma^\text{gen}_\infty \rangle_\text{min}
  & = \kB \ln 2 \lim_{N \rightarrow \infty}
  (\I[Y'_{0:N};X_N] \! - \! \I[Y'_{0:N};X_{N+1}, Y'_N]) \\
  & = \tfrac{2}{3} \kB \ln 2 \\
  & \approx 0.462 ~\kB
  ~.
\end{align*}
With every time step, this predictive ratchet stores information about its
past, but it also erases information, dissipating $2/3$ of a bit worth of
correlations without leveraging them. Those correlations could have been used
to reverse the process if they had been turned into work. They are used by the
retrodictive ratchet, though, which stores just enough information about its
past to generate the future.

It was previously shown that storing unnecessary information about the past
leads to additional transient dissipation when generating a pattern
\cite{Boyd16e,Garn15}. This cost also arises from implementation. However, our
measure of modularity dissipation shows that there are implementation costs
that persist through time. The two locally-operating generators of the Golden
Mean Process perform the same computation, but have different bounds on their
dissipation per time step. Thus, the additional work investment required to
generate the process grows linearly with time for the \eM\ implementation, but
is zero for the retrodictive implementation.

\begin{figure}[tbp]
\centering
\includegraphics[width= \columnwidth]{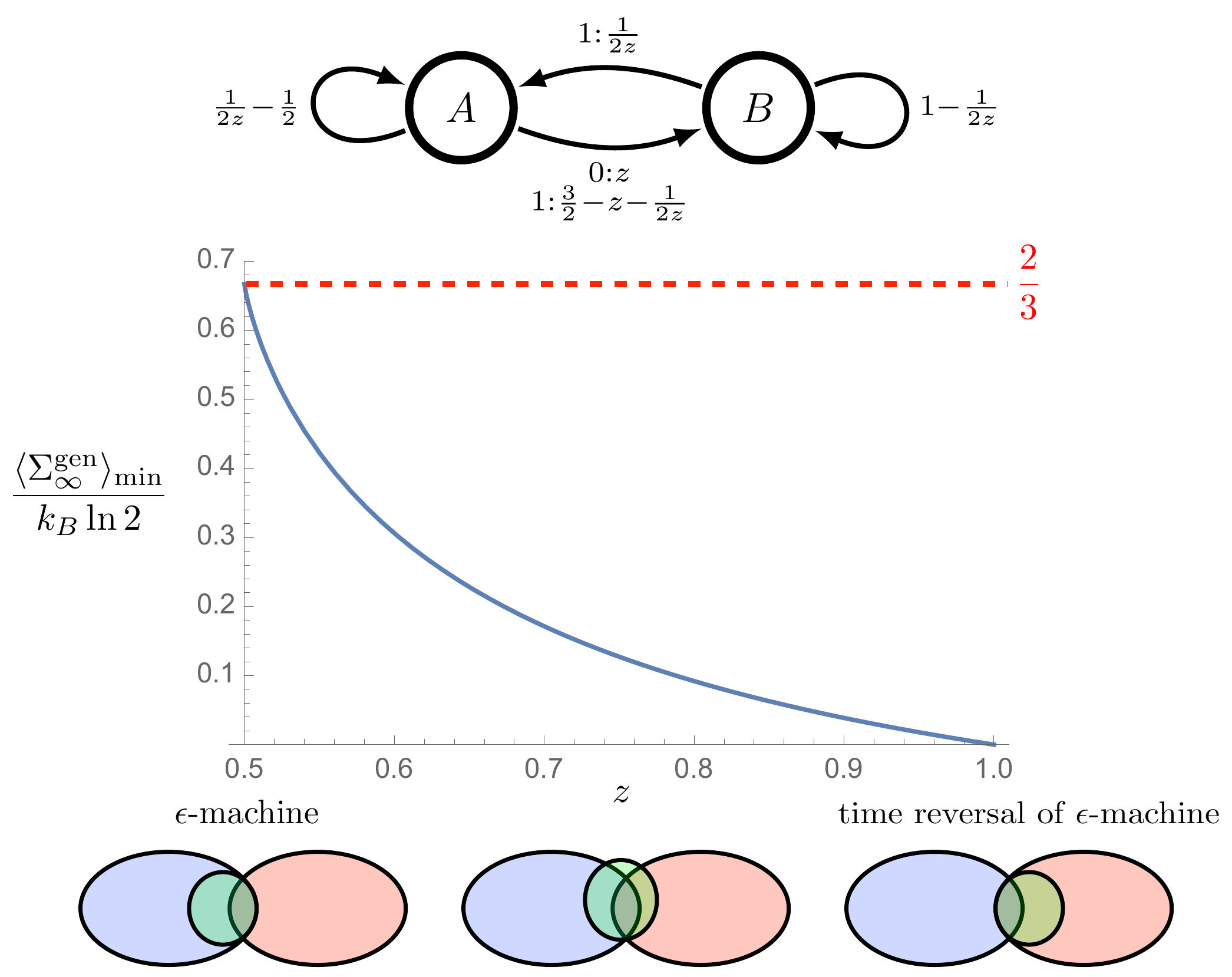}
\caption{(Top) A parametrized family of HMMs that generate the Golden Mean
	Process for $z \in [.5,1]$. (Middle) As parameter $z$ increases, the
	information stored in the hidden states about the output past decreases.
	At $z=0.5$ the HMM is the \eM, whose states are a function of the past. At
	$z=1.0$, the HMM is the time reversal of the reverse-time \eM, whose states
	are a function of the future. The modularity dissipation decreases
	monotonically as $z$ increases and the hidden states' memory of the past
	decreases. (Bottom) Information diagrams corresponding to the end cases and
	a middle case. Labeling as in Fig. \ref{fig:PredictiveVsRetrodictive}.
	}
\label{fig:zModel}
\end{figure} 

Moreover, we can consider generators that fall in-between these extremes using
the parametrized HMM shown in Fig. \ref{fig:zModel} (top). This HMM,
parametrized by $z$, produces the Golden Mean Process at all $z \in [.5,1]$,
but the hidden states share less and less information with the output past as
$z$ increases, as shown by Ref. \cite{Elli11a}. One extreme $z=0.5$ corresponds
to the minimal predictive generator, the \eM. The other at $z=1$ corresponds to
the minimal retrodictive generator, the time reversal of the reverse-time \eM.
The graph there plots the modularity dissipation as a function of $z$. It
decreases with $z$, suggesting that the unnecessary memory of the past leads to
additional dissipation. So, while we have only proved that retrodictive
generators are maximally efficient, this demonstrates that extending beyond
that class can lead to unnecessary dissipation and that there may be a direct
relationship between unnecessary memory and dissipation.

Taken altogether, we see that the thermodynamic consequences of localized
information processing lead to direct principles for efficient information
transduction. Analyzing the most general case of transducing arbitrary
structured processes into other arbitrary structured processes remains a
challenge. That said, pattern generators and pattern extractors have elegantly
symmetric conditions for efficiency that give insight into the range of
possibilities. Pattern generators are effectively the time-reversal of pattern
extractors, which turn structured inputs into structureless outputs. As such
they are most efficient when retrodictive, which is the time-reversal of being
predictive. Figure \ref{fig:PredictiveVsRetrodictive} illustrated graphically
how the predictive \eM\ captures past correlations and stores the necessary
information about the past, while the retrodictive ratchet's states are
analogous, but store information about the future instead. This may seem
unphysical---as if the ratchet is anticipating the future. However, since the
ratchet generates the output future, this anticipation is entirely physical,
because the ratchet controls the future, as opposed to mysteriously predicting
it, as an oracle would.

\section{Conclusion}

Modularity is a key design theme in physical information processing, since it
gives the flexibility to stitch together many elementary logical operations to
implement a much larger computation. Any classical computation can be composed
from local operations on a subset of information reservoir observables.
Modularity is also key to biological organization, its functioning, and our
understanding of these \cite{Alon07a}.

However, there is an irretrievable thermodynamic cost, the modularity
dissipation,  to this localized computing, which we quantified in terms of the
global entropy production. This modularity-induced entropy production is
proportional to the reduction of global correlations between the local and
interacting portion of the information reservoir and the fixed, noninteracting
portion. This measure forms the basis for designing thermodynamically efficient
information processing. It is proportional to the additional work investment
required by the modular form of the computation, beyond the work required by a
globally integrated and reversible computation.

Turing machine-like information ratchets provide a natural application for this
new measure of efficient information processing, since they process information
in a symbol string through a sequence of local operations. The modularity
dissipation allows us to determine which implementations are able to achieve
the asymptotic bound set by the IPSL which, substantially generalizing
Landauer's bound, says that any type of structure in the input can be used as a
thermal resource and any structure in the output has a thermodynamic cost.
There are many different ratchet implementations that perform a given
computation, in that they map inputs to outputs in the same way. However, if we
want an implementation to be thermodynamically efficient, the modularity
dissipation, monitored by the global entropy production, must be minimized.
Conversely, we now appreciate why there are many implementations that dissipate
and are thus irreversible. This establishes modularity dissipation as a new
thermodynamic cost, due purely to an implementation's architecture, that
complements Landauer's bound on isolated logical operations.

We noted that there are not yet general principles for designing devices that
minimize modularity dissipation and thus work investment for arbitrary
information transduction. However, for the particular cases of pattern
generation and pattern extraction we find that there are prescribed classes of
ratchets that are guaranteed to be dissipationless, if operated
quasistatically. The ratchet states of these devices are able to store and
leverage the global correlations among the symbol strings, which means that it
is possible to achieve the reversibility of globally integrated information
processing but with modular computational design. Thus, while modular
computation often results in dissipating global correlations, this inefficiency
can be avoided when designing processors by employing the tools of computations
mechanics outlined here.

\section*{Acknowledgments}

As an External Faculty member, JPC thanks the Santa Fe Institute for its
hospitality during visits. This material is based upon work supported by, or in
part by, John Templeton Foundation grant 52095, Foundational Questions
Institute grant FQXi-RFP-1609, and the U. S. Army Research Laboratory and the
U. S. Army Research Office under contract W911NF-13-1-0390.

\appendix

\section{Quasistatic Markov Channels}
\label{app:Quasistatic Markov Channels}

To satisfy information-theoretic bounds on work dissipation, we describe a
quasistatic channel where we slowly change system energies to manipulate the
distribution over $\ISystem$'s states. Precisely, our challenge is to evolve
over time interval $(t,t+\tau)$ an input distribution $\Pr(\RVSystem_t =
\rvsystem_t)$ according the Markov channel $M$, so that system's conditional
probability at time $t+\tau$ is:
\begin{align*}
\Pr(\RVSystem_{t+\tau} = \rvsystem_{t+\tau} | \RVSystem_t = \rvsystem_t)
  = M_{\rvsystem_t \rightarrow \rvsystem_{t+\tau}}
  ~.
\end{align*}
Making this as efficient as possible in a thermal environment at temperature
$T$ means ensuring that the work invested in the evolution achieves the lower
bound:
\begin{align*}
\langle W \rangle \geq k_B T \ln 2 ( \H[\RVSystem_t] - \H[\RVSystem_{t+\tau}])
  ~.
\end{align*}
This expresses the Second Law of Thermodynamics for the system in contact with a
heat bath.

To ensure the appropriate conditional distribution, we introduce an ancillary
system $\ISystem'$, which is a copy of $\ISystem$. So that it is efficient, we
take $\tau$ to be large with respect to the system's relaxation time scale and
break the overall process into three steps that occur over the time intervals
$(t,t+ \tau_0)$, $(t +\tau_0, t+\tau_1)$, and $(t+\tau_1,t+\tau)$, where
$0<\tau_0<\tau_1<\tau$.

Our method of manipulating $\ISystem$ and $\ISystem'$ is to control the energy
$E(t,\rvsystem,\rvsystem')$ of the joint state $\rvsystem \otimes \rvsystem'
\in \ISystem \otimes \ISystem'$ at time $t$. We also control whether or not
probability is allowed to flow in $\ISystem$ or $\ISystem'$. This corresponds
to raising or lowering energy barriers between system states.

At the beginning of the control protocol we choose $\ISystem'$ to be in a
uniform distribution uncorrelated with $\ISystem$. This means the joint
distribution can be expressed:
\begin{align}
\Pr(\RVSystem_t=\rvsystem_t,\RVSystem'_t=\rvsystem_t')
  = \frac{\Pr(\RVSystem_t=\rvsystem_t)}{|\ISystem'|}
  ~.
\end{align}  
Since we are manipulating an energetically mute information reservoir, we also
start with the system in a uniformly zero-energy state over the joint states of
$\ISystem$ and $\ISystem'$:
\begin{align}
E(t,z,z')=0
  ~.
\end{align}
While this energy and the distribution change when executing the protocol, we
return $\ISystem'$ to the independent uniform distribution and the energy to
zero at the end of the protocol. This ensures consistency and modularity.
However, the same results can be achieved by choosing other starting energies
with $\ISystem'$ in other distributions.

The three steps that evolve this system to quasistatically implement the Markov
channel $M$ are as follows:
\begin{enumerate}
\item Over the time interval $(t,t+\tau_0)$, continuously change the energy
	such that the energy at the end of the interval $E(t+\tau_0,z,z')$ obeys
	the relation:
\begin{align*}
e^{-(E(t+\tau_0,\rvsystem,\rvsystem') - F(t+\tau_0))/k_B T}
  = \Pr(\RVSystem_t=\rvsystem) M_{\rvsystem \rightarrow \rvsystem'}
  ~,
\end{align*}
while allowing state space and probability to flow in $\ISystem'$, but not in
$\ISystem$. Since the protocol is quasistatic, $\ISystem'$ follows the
Boltzmann distribution and at time $t+\tau_0$ the distribution over
$\ISystem \otimes \ISystem'$ is:
\begin{align*}
\Pr(\RVSystem_{t+\tau_0}=\rvsystem,\RVSystem'_{t+\tau_0}=\rvsystem')
  = \Pr(\RVSystem_t=\rvsystem)
  M_{\rvsystem \rightarrow \rvsystem'}
  ~.
\end{align*}
This yields the conditional distribution of the current ancillary variable
$\RVSystem'_{t+\tau}$ on the initial system variable $\RVSystem_t$:
\begin{align*}
\Pr(\RVSystem'_{t+\tau_0}=\rvsystem'|\RVSystem_t=\rvsystem)
  = M_{\rvsystem \rightarrow \rvsystem'}
  ~,
\end{align*}
since the system variable $\RVSystem_t$ remains fixed over the interval. This
protocol effectively applies the Markov channel $M$ to evolve from $\ISystem$
to $\ISystem'$. However, we want the Markov channel to apply strictly to
$\ISystem$.

Being a quasistatic protocol, there is no entropy production and the work
flow is simply the change in nonequilibrium free energy:
\begin{align*}
\langle W_1 \rangle & = \Delta F^\text{neq} \\
  & = \Delta \langle E \rangle - T \Delta S[\RVSystem , \RVSystem']
  ~.
\end{align*}
Since the average initial energy is uniformly zero, the change in average
energy is the average energy at time $t+\tau_0$. And so, we can express
the work done:
\begin{align*}
\langle W_1 \rangle
  & =  \langle E(t+\tau_0) \rangle -T \Delta S[\RVSystem , \RVSystem'] \\
  & =  \langle E(t+\tau_0) \rangle \\
  & \qquad + \kB T \ln 2 (\H[\RVSystem_t,\RVSystem'_t]
  -\H[\RVSystem_{t+\tau_0},\RVSystem_{t+\tau_0}'])
  ~.
\end{align*}

\item Now, swap the states of $\ISystem$ and $\ISystem'$ over the time interval $(t+\tau_0,t+\tau_1)$. This is logically reversible.
Thus, it can be done without any work investment over the second time interval:
\begin{align}
\langle W_2 \rangle=0
  ~.
\end{align}
The result is that the energies and probability distributions are flipped with
regard to exchange of the system $\ISystem$ and ancillary system $\ISystem'$:
\begin{align*}
E(t+\tau_1,\rvsystem,\rvsystem')
  & = E(t+\tau_0,\rvsystem',\rvsystem) \\
  \Pr(\RVSystem_{t+\tau_1} \! = \! \rvsystem,
  \RVSystem'_{t+\tau_1}\! =\! \rvsystem')
  & =\Pr(\RVSystem_{t+\tau_0}\! =\! \rvsystem',\RVSystem'_{t+\tau_0}\!
  =\! \rvsystem)
  ~.
\end{align*}
Most importantly, however, this means that the conditional probability of the
current system variable is given by $M$:
\begin{align*}
Pr(\RVSystem_{t+\tau_1}=\rvsystem'|\RVSystem_t=\rvsystem)
  & = \Pr(\RVSystem'_{t+\tau_0}=\rvsystem'|\RVSystem_t=\rvsystem) \\
  & = M_{\rvsystem \rightarrow \rvsystem'}
  ~.
\end{align*}
The ancillary system must still be reset to a uniform and uncorrelated state and the energies must be reset.

\item Finally, we again hold $\ISystem$'s state fixed while allowing
$\ISystem'$ to change over the time interval $(t+\tau_1,t+\tau)$ as we change
the energy, ending at $E(t+\tau,\rvsystem,\rvsystem')=0$.  This quasistatically
brings the joint distribution to one where the ancillary system is uniform and
independent of $\ISystem$:
\begin{align}
\Pr(\RVSystem_{t+\tau}=\rvsystem,\RVSystem_{t+\tau}'=\rvsystem')
  = \frac{\Pr(\RVSystem_{t+\tau}=\rvsystem)}{|\mathcal{\RVSystem}'|}
  ~.
\end{align}
Again, the invested work is the change in average energy plus the change in
thermodynamic entropy of the joint system:
\begin{align*}
\langle W_3 \rangle & =\langle \Delta E \rangle \\
  & \quad + \kB T \ln 2
  (\H[\RVSystem_{t+\tau_1},\RVSystem'_{t+\tau_1}]
  - \H[\RVSystem_{t+\tau},\RVSystem_{t+\tau}']) \\
  & = -\langle E(t+\tau_1) \rangle \\
  & \quad + \kB T \ln 2
  (\H[\RVSystem_{t+\tau_1},\RVSystem'_{t+\tau_1}]
  - \H[\RVSystem_{t+\tau},\RVSystem_{t+\tau}'])
  ~.
\end{align*}
This ends this three-step protocol.
\end{enumerate}

Summing up the heat terms, gives the total dissipation:
\begin{align*}
\langle W_\text{total} \rangle
  & = \langle W_t \rangle +\langle W_2 \rangle +\langle W_3 \rangle \\
  & = \kB T \ln2
  (\H[\RVSystem_t,\RVSystem'_t]
  -\H[\RVSystem_{t+\tau_0},\RVSystem_{t+\tau_0}']) \\
  & \quad + \kB
  T(\H[\RVSystem_{t+\tau_1},\RVSystem'_{t+\tau_1}]
  - \H[\RVSystem_{t+\tau},\RVSystem_{t+\tau}']) \\
  & \quad + \langle E(t+\tau_0) \rangle-\langle E(t+\tau_1) \rangle
  ~.
\end{align*}

Recall that the distributions $\Pr(\RVSystem_{t+\tau_1},\RVSystem'_{t+\tau_1})$
and $\Pr(\RVSystem_{t+\tau_0},\RVSystem_{t+\tau_0}')$, as well as
$E(t+\tau_0,\rvsystem,\rvsystem')$ and $E(t+\tau_1,\rvsystem,\rvsystem')$, are
identical under exchange of $\ISystem$ and $\ISystem'$, so
$H[\RVSystem_{t+\tau_1},\RVSystem'_{t+\tau_1}]=H[\RVSystem_{t+\tau_0},\RVSystem_{t+\tau_0}']$
and $\langle E(t+\tau_0) \rangle=\langle E(t+\tau_1) \rangle$. Additionally, we
know that both the starting and ending distributions have a uniform and
uncorrelated ancillary system, so their entropies can be expressed:
\begin{align}
\H[\RVSystem_t,\RVSystem'_t] & = \H[\RVSystem_t]+\log_2 |\ISystem'| \\
  \H[\RVSystem_{t+\tau},\RVSystem'_{t+\tau}]
  & = \H[\RVSystem_{t+\tau}]+\log_2 |\ISystem'|
  ~.
\end{align}
Substituting this in to the above expression for total invested work, we find
that we achieve the lower bound with this protocol:
\begin{align}
\langle W_\text{total} \rangle
  = \kB T \ln 2 (\H[\RVSystem_t] - \H[\RVSystem_{t+\tau}])
  ~.
\end{align}
Thus, the protocol implements a Markov channel that achieves the
information-theoretic bounds. It is similar to that described in Ref.
\cite{Garn15}.  

The basic principle underlying the protocol is that when manipulating system
energies to change state space, choose the energies so that there is no
instantaneous probability flow. That is, if one stops changing the energies,
the distribution will not change. However, there are cases in which it is
impossible prevent instantaneous flow. Then, there are necessarily
inefficiencies that arise from the dissipation of the distribution flowing out
of equilibrium.

\section{Transducer Dissipation}
\label{app:Dissipation of Predictive and Retrodictive Transducers}

\subsection{Predictive Extractors}

For a pattern extractor, being reversible means that the transducer is
predictive of the input process. More precisely, an extracting transducer
that produces zero entropy is equivalent to it being a predictor of its input.

As discussed earlier, a reversible extractor satisfies:
\begin{align*}
\I[Y_{N+1:\infty};X_{N+1}] = \I[Y_{N+1:\infty};X_{N}Y_N]
  ~,
\end{align*}
for all $N$, since it must be reversible at every step to be fully reversible.
The physical ratchet being predictive of the input means two things. It means
that $X_N$ shields the past $Y_{0:N}$ from the future $Y_{N:\infty}$. This is
equivalent to the mutual information between the past and future vanishing when
conditioned on the ratchet state:
\begin{align*}
\I[Y_{0:N};Y_{N:\infty}|X_N] = 0
  ~.
\end{align*}
Note that this also implies that any subset of the past or future is
independent of any other subset conditioned on the ratchet state:
\begin{align*}
\I[Y_{a:b};Y_{c:d}|X_N] = 0 \text{ where } b \leq N \text{ and } c \geq N
  ~.
\end{align*}
The other feature of a predictive transducer is that the past shields the
ratchet state from the future:
\begin{align*}
\I[X_N;Y_{N:\infty}|Y_{0:N}] = 0
  ~.
\end{align*}
This is guaranteed by the fact that transducers are nonanticipatory: they
cannot predict future inputs outside of their correlations with past inputs.

\begin{figure}[tbp]
\centering
\includegraphics[width=.9\columnwidth]{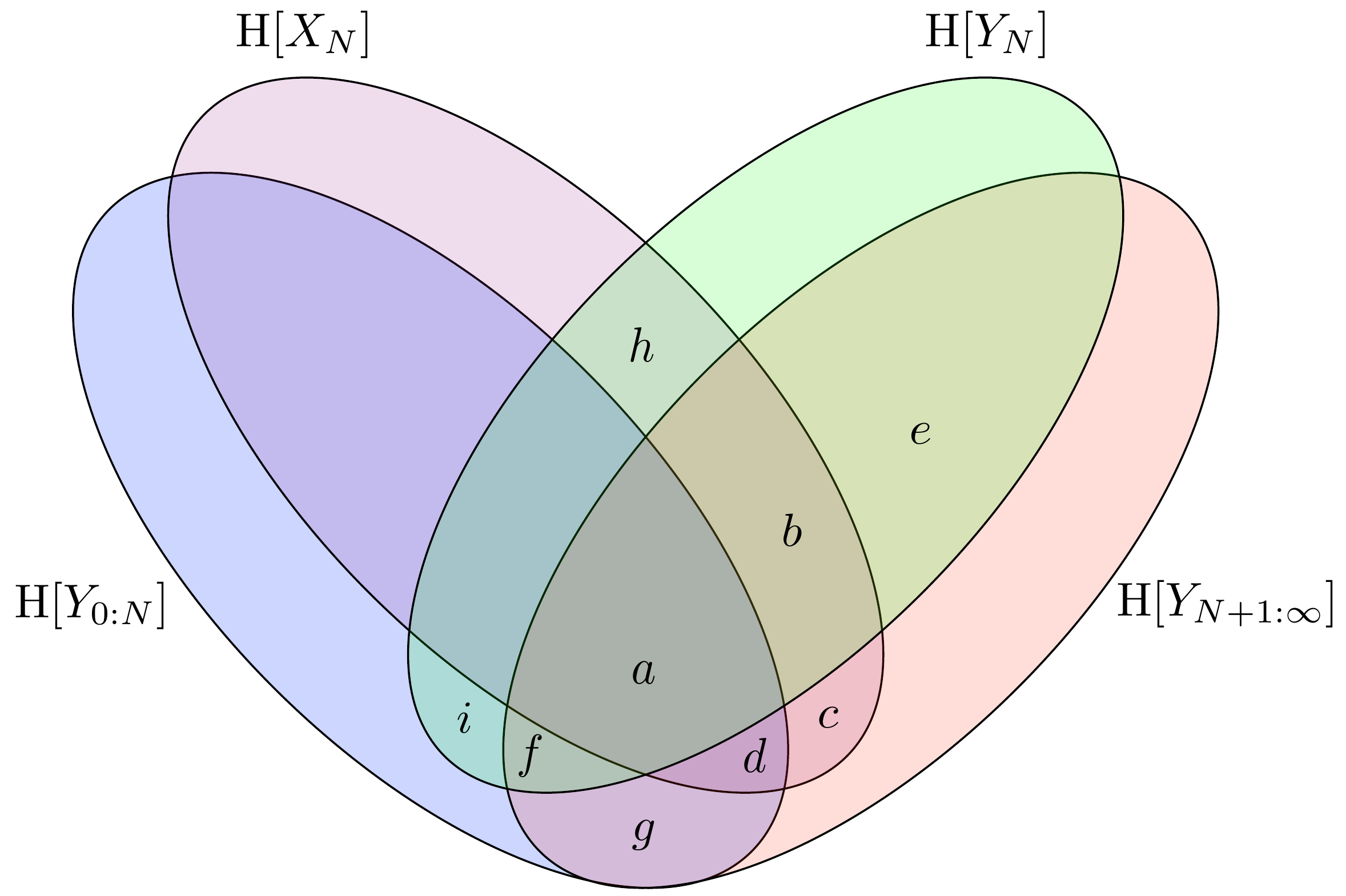}
\caption{Information diagram for dependencies between the input past $Y_{0:N}$,
	next input $Y_N$, current ratchet state $X_N$, and input future
	$Y_{N+1:\infty}$, excluding the next input. We label certain information
	atoms to help illustrate the algebraic steps in the associated proof.
 	}
\label{fig:EfficientPredictor}
\end{figure}

We start by showing that if the ratchet is predictive, then the entropy
production vanishes. It is useful to note that being predictive is equivalent to
being as anticipatory as possible and having:
\begin{align*}
\I[X_N;Y_{N:\infty}] = \I[Y_{0:N};Y_{N:\infty}]
  ~,
\end{align*}
which can be seen by subtracting $\I[Y_{0:N};Y_{N:\infty};X_N]$ from each side
of the immediately preceding expression.
Thus, it is sufficient to show that the mutual information between the partial
input future $Y_{N+1:\infty}$ and the joint distribution of the predictive
variable $X_N$ and next output $Y_N$ is the same as mutual information with the
joint variable $(Y_{0:N}, Y_N) = Y_{0:N+1}$ of the past inputs and the next
input:
\begin{align*}
\I[Y_{N+1:\infty};X_{N}, Y_N] = \I[Y_{N+1:\infty};Y_{0:N}, Y_N]
  ~.
\end{align*}
To show this for a predictive variable, we use Fig.
\ref{fig:EfficientPredictor}, which displays the information diagram for all
four variables with the information atoms of interest labeled.

Assuming that $X_N$ is predictive zeros out a number of information atoms, as
shown below:
\begin{align*}
\I[X_N;Y_N,Y_{N+1:\infty}|Y_{0:N}] & = b+c+h = 0 \\
  \I[X_N;Y_N|Y_{0:N}] & = b+h = 0 \\
  \I[Y_{0:N};Y_N,Y_{N+1:\infty}|X_N] & = i+f+g = 0 \\
  \I[Y_{0:N};Y_N|X_N] & = i+f = 0
  ~.
\end{align*}
These four equations make it clear that $g=c=0$. Thus, substituting
$\I[Y_{N+1:\infty};X_{N},Y_N]=a+b+c+d+e+f$ and
$\I[Y_{N+1:\infty};Y_{0:N},Y_N]=a+b+d+e+f+g$, we find that their difference
vanishes:
\begin{align*}
\I[Y_{N+1:\infty};X_{N},Y_N]-I[Y_{N+1:\infty};Y_{0:N},Y_N] & = c-g \\
  & = 0
  ~.
\end{align*}

There is zero dissipation, since $X_{N+1}$ is also predictive, meaning
$\I[Y_{N+1:\infty};Y_{0:N},Y_N] = \I[Y_{N+1:\infty};X_{N+1}]$, so:
\begin{align*}
\frac{\langle \Sigma^\text{ext}_N \rangle_\text{min}}{\kB T \ln 2}
  & = \I[Y_{N+1:\infty};X_{N},Y_N]-I[Y_{N+1:\infty};X_{N+1}] \\
  & = \I[Y_{N+1:\infty};X_{N},Y_N]-I[Y_{N+1:\infty};Y_{0:N+1}] \\
  & = 0
  ~.
\end{align*}

Going the other direction, using zero entropy production to prove that $X_N$ is
predictive for all $N$ is now simple.

We already showed that
$\I[Y_{N+1:\infty};X_{N},Y_N]=\I[Y_{N+1:\infty};Y_{0:N},Y_N]$ if $X_N$ is
predictive. Combining with zero entropy production
($\I[Y_{N+1:\infty};X_{N+1}]=\I[Y_{N+1:\infty};X_{N},Y_N]$) immediately implies
that $X_{N+1}$ is predictive, since
$\I[Y_{N+1:\infty};X_{N+1}]=\I[Y_{N+1:\infty};Y_{0:N},Y_N]$ plus the fact that
$X_{N+1}$ is equivalent to $X_{N+1}$ being predictive.

With this recursive relation, all that is left to establish is the base case,
that $X_0$ is predictive. Applying zero entropy production again we find the
relation necessary for prediction:
\begin{align*}
\I[Y_{1:\infty};X_{1}] & = \I[Y_{1:\infty};X_{0},Y_0] \\
  & = \I[Y_{1:\infty};Y_0]
  ~,
\end{align*}
From this, we find the equivalence
$\I[Y_{1:\infty};Y_0]=\I[Y_{1:\infty};X_{0},Y_0]$, since $X_0$ is independent of
all inputs, due to it being nonanticipatory. Thus, zero entropy production is
equivalent to predictive ratchets for pattern extractors.

\subsection{Retrodictive Generators}

An analogous argument can be made to show the relationship between retrodiction
and zero entropy production for pattern generators, which are essentially time
reversed extractors.

\begin{figure}[tbp]
\centering
\includegraphics[width=.9\columnwidth]{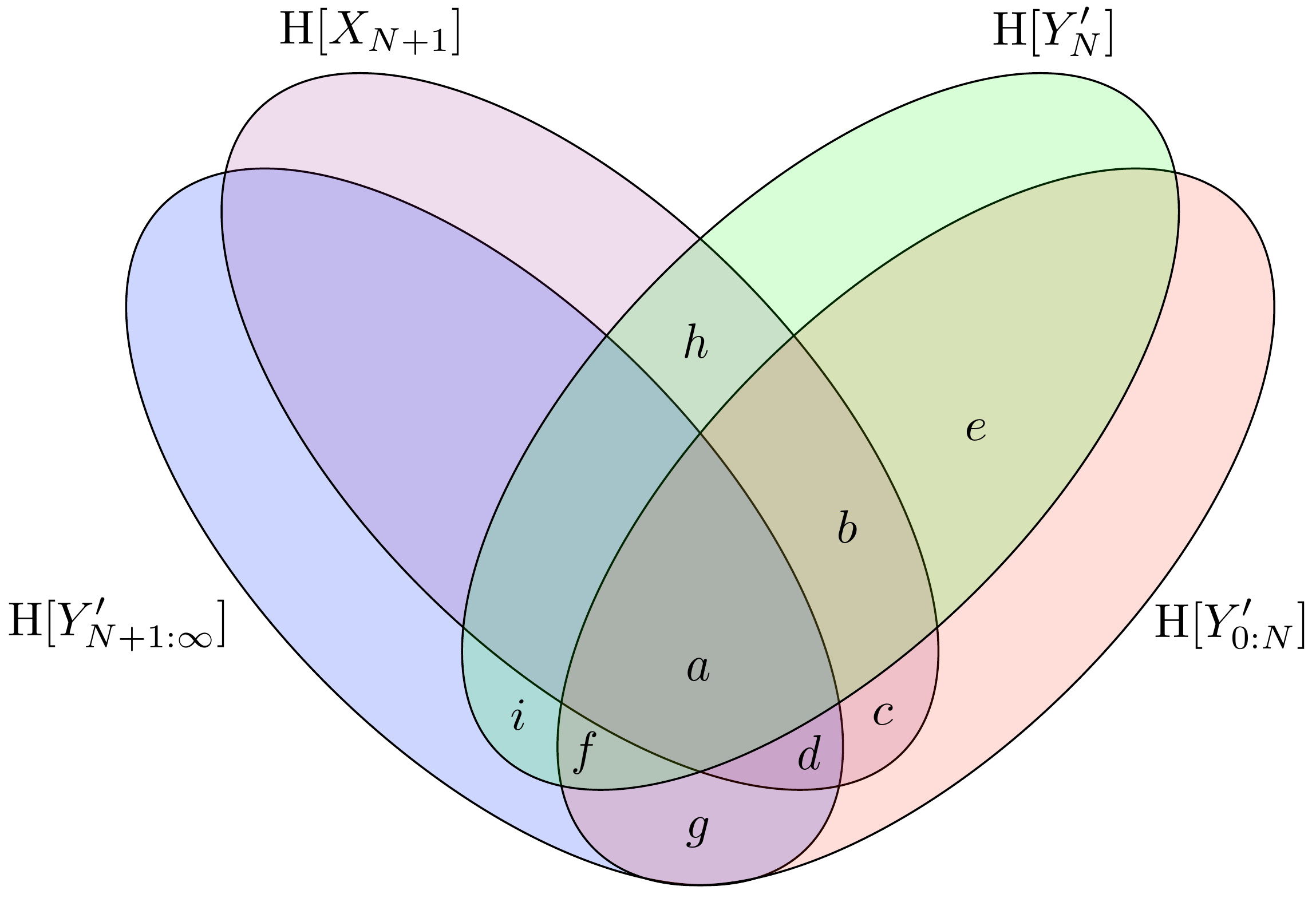}
\caption{Information shared between the output past $Y'_{0:N}$, next output
	$Y'_N$, next ratchet state $X_{N+1}$, and output future $Y'_{N+1:\infty}$,
	excluding the next input. Key information atoms are labeled.
	}
\label{fig:EfficientGenerator}
\end{figure}

Efficient pattern generators must satisfy:
\begin{align*}
\I[Y_{0:N}';X_{N}] = \I[Y_{0:N}';X_{N+1} Y_N']
  ~.
\end{align*}
The ratchet being retrodictive means that the ratchet state $X_N$ shields the
past $Y'_{0:N}$ from the future $Y'_{N:\infty}$ and that the future shields the
ratchet from the past:
\begin{align*}
\I[Y'_{0:N};Y'_{N:\infty}|X_N] & = 0 \\
   \I[Y'_{0:N};X_N | Y'_{N:\infty}] & = 0
   ~.
\end{align*}
Note that generators necessarily shield past from future
$\I[Y'_{0:N};Y'_{N:\infty}|X_N] = 0$, since all temporal correlations must be
stored in the generator's states. Thus, for a generator, being retrodictive is
equivalent to:
\begin{align*}
\I[Y'_{0:N};X_N] = \I[Y'_{0:N};Y'_{N:\infty}]
  ~.
\end{align*}
This can be seen by subtracting $\I[Y'_{0:N};X_N ; Y'_{N:\infty}]$ from both
sides, much as done with the extractor.

First, to show that being retrodictive implies zero minimal entropy production,
it is sufficient to show that:
\begin{align*}
\I[Y_{0:N}';X_{N+1}, Y_N'] = \I[Y_{0:N}'; Y_{N:\infty}']
  ~,
\end{align*}
since we know that $\I[Y_{0:N}';X_{N}]=\I[Y_{0:N}'; Y_{N:\infty}']$. To do
this, consider the information diagram in Fig. \ref{fig:EfficientGenerator}.
There we see that the difference between the two mutual informations of
interest reduce to the difference between the two information atoms:
\begin{align*}
\I[Y_{0:N}';X_{N+1} Y_N'] - \I[Y_{0:N}'; Y_{N:\infty}'] = c-g
  ~.
\end{align*}
The fact that the ratchet state $X_{N+1}$ shields the past $Y'_{0:N+1}$ from
the future $Y'_{N+1:\infty}$ and the future shields the ratchet from the past
gives us the following four relations:
\begin{align*}
\I[Y'_{0:N}Y'_N;Y'_{N+1:\infty}|X_{N+1}] & = i+f+g = 0 \\
  \I[Y'_N;Y'_{N+1:\infty}|X_{N+1}] & = i+f = 0 \\
  \I[Y'_{0:N}Y'_N;X_{N+1}|Y'_{N+1:\infty}] & = h+b+c = 0 \\
  \I[Y'_N;X_{N+1}|Y'_{N+1:\infty}] & = h+b = 0
  ~.
\end{align*}
These equations show that that $c=g=0$ and thus:
\begin{align*}
\frac{\langle \Sigma^\text{gen}_N \rangle_\text{min}}{\kB T \ln 2} = 0
  ~.
\end{align*}

Going the other direction---zero entropy production implies
retrodiction---requires that we use
$\I[Y_{0:N}';X_{N}]=\I[Y_{0:N}';X_{N+1},Y_N']$ to show $\I[Y'_{0:N};X_N]
=\I[Y'_{0:N};Y'_{N:\infty}]$. If $X_{N+1}$ is retrodictive, then we can show
that $X_N$ must be as well. However, this makes the base case of the recursion
difficult, since there is not yet a reason to conclude that $X_\infty$ is
retrodictive. While we conjecture the equivalence of optimally retrodictive
generators and efficient generators, at this point we can only conclusively say
that retrodictive generators are also efficient.

\bibliography{chaos}

\end{document}

%% file: cmechabbrev.tex
\newcommand{\eM}     {\mbox{$\epsilon$-machine}}

\newcommand{\CausalState}   { \mathcal{S} }

\newcommand{\hmu}       {h_\mu}

\newcommand{\forward}{+}
\newcommand{\reverse}{-}
\newcommand{\forwardreverse}{\pm} 

\newcommand{\FutureCausalState} { {\CausalState}^{\forward} }

\newcommand{\PastCausalState}   { {\CausalState}^{\reverse} }

\newcommand{\lastindex}[2]{
  \edef\tempa{0}
  \edef\tempb{#2}
  \ifx\tempa\tempb
    \edef\tempc{#1}
  \else
    \edef\tempa{0}
    \edef\tempb{#1}
    \ifx\tempa\tempb
      \edef\tempc{#2}
    \else
      \edef\tempc{#1+#2}
    \fi
  \fi
  \tempc
}

\newcommand{\I}{\mathbf{I}}

\newcommand{\CSjoint}[1][,]{
   \edef\tempa{:}
   \edef\tempb{#1}
   \ifx\tempa\tempb
      \ensuremath{\FutureCausalState\!#1\PastCausalState}
   \else
      \ensuremath{\FutureCausalState#1\PastCausalState}
   \fi
}

\newif\ifpm
\edef\tempa{\forwardreverse}
\edef\tempb{\pm}
\ifx\tempa\tempb
   \pmfalse
\else
   \pmtrue
\fi